\documentclass[prd,superscriptaddress,amsfonts,amssymb,amsmath,showpacs,twocolumn]{revtex4-2}
\usepackage{bm}
\usepackage{amsfonts}
\usepackage{latexsym}
\usepackage[latin1]{inputenc}
\usepackage{graphicx}
\usepackage{amsmath}
\usepackage{palatino}
\usepackage{mathpazo}
\usepackage{textcomp}
\usepackage{float}
\usepackage{booktabs}
\usepackage{dcolumn}
\usepackage{ragged2e}
\usepackage{hyperref}
\hypersetup{colorlinks,citecolor=blue}
\hypersetup{colorlinks=true,linkcolor=red,filecolor=magenta,    urlcolor=blue}
\usepackage{amsmath}
\usepackage{xcolor}
\usepackage{orcidlink}
\usepackage[caption=false]{subfig}
\usepackage{commath}
\def\jnl@style{\it}
\def\aaref@jnl#1{{\jnl@style#1}}

\def\aaref@jnl#1{{\jnl@style#1}}

\def\aj{\aaref@jnl{AJ}}                   
\def\apj{\aaref@jnl{ApJ}}                 
\def\apjl{\aaref@jnl{ApJ}}                
\def\apjs{\aaref@jnl{ApJS}}               
\def\apss{\aaref@jnl{Ap\&SS}}             
\def\aap{\aaref@jnl{A\&A}}                
\def\aapr{\aaref@jnl{A\&A~Rev.}}          
\def\aaps{\aaref@jnl{A\&AS}}              
\def\mnras{\aaref@jnl{Mon.~Not.~Roy.~Astron.~Soc.}}             
\def\prd{\aaref@jnl{Phys.~Rev.~D}}        
\def\plb{\aaref@jnl{Phys.~Lett.~B}}        
\def\prc{\aaref@jnl{Phys.~Rev.~C}}  
\def\prl{\aaref@jnl{Phys.~Rev.~Lett.}}    
\def\qjras{\aaref@jnl{QJRAS}}             
\def\skytel{\aaref@jnl{S\&T}}             
\def\ssr{\aaref@jnl{Space~Sci.~Rev.}}     
\def\zap{\aaref@jnl{ZAp}}                 
\def\nat{\aaref@jnl{Nature}}              
\def\aplett{\aaref@jnl{Astrophys.~Lett.}} 
\def\apspr{\aaref@jnl{Astrophys.~Space~Phys.~Res.}} 
\def\physrep{\aaref@jnl{Phys.~Rep.}}      
\def\physscr{\aaref@jnl{Phys.~Scr}}       
\def\commat{\aaref@jnl{Comm.~Math.~Phys.}}              
\def\science{\aaref@jnl{Science}}               
\def\cqg{\aaref@jnl{Classical Quant.~Grav.}}            
\def\jpcs{\aaref@jnl{JPCS}}                                     
\def\ijmpd{\aaref@jnl{Int.~J.~Mod.~Phys.~D}}                    
\def\grg{\aaref@jnl{Gen.~Relat.~Gravit.}}               
\def\rpp{\aaref@jnl{Rep.~Prog.~Phys.}}          
\def\npa{\aaref@jnl{Nucl.~Phys.~A}}        
\def\lrr{\aaref@jnl{Living Rev.~Rel.}}                   
\def\jcap{\aaref@jnl{J.~Cosmology Astropart.~Phys.}}    
\def\rmp{\aaref@jnl{Rev.~Mod.~Phys.}}   
\def\epjc{\aaref@jnl{Eur.~Phys.~J.~C}}

\allowdisplaybreaks[1]

\begin{document}
\color{black}       
\title{The effect of pressure anisotropy on quark stars structure in the Starobinsky model}

\author{Takol Tangphati
\orcidlink{0000-0002-6818-8404}} 
\email{takoltang@gmail.com}
\affiliation{School of Science, Walailak University, Thasala, \\Nakhon Si Thammarat, 80160, Thailand}
\affiliation{Research Center for Theoretical Simulation and Applied Research in Bioscience and Sensing, Walailak University, Thasala, Nakhon Si Thammarat 80160, Thailand}

\author{\.{I}zzet Sakall{\i} \orcidlink{0000-0001-7827-9476}}
\email{izzet.sakalli@emu.edu.tr}
\affiliation {Physics Department, Eastern Mediterranean University, Famagusta 99628, North Cyprus via Mersin 10, Turkey.}

\author{Ayan Banerjee \orcidlink{0000-0003-3422-8233}} 
\email{ayanbanerjeemath@gmail.com}
\affiliation{Astrophysics and Cosmology Research Unit, School of Mathematics, Statistics and Computer Science, University of KwaZulu--Natal, Private Bag X54001, Durban 4000, South Africa}

\author{Anirudh Pradhan
\orcidlink{0000-0002-1932-8431}} 
\email[]{pradhan.anirudh@gmail.com}
\affiliation{Centre for Cosmology, Astrophysics and Space Science, GLA University, Mathura-281 406, Uttar Pradesh, India}


\date{\today}

\begin{abstract}
The structure and stability of quark stars (QSs) made of interacting quark matter are discussed in this study, taking color superconductivity and perturbative QCD corrections into account. By combining this EoS with the Tolman-Oppenheimer-Volkoff (TOV) equations, we explore the mass-radius ($M-R$) relations of QSs. The analysis is conducted within the framework of $R^2$ gravity, where the gravity model is described by $f(R) = R + a R^2$. Our primary goal is to investigate how variations in the $R^2$ gravity parameter $a$ affect the mass-radius and mass-central density ($M-\rho_c$) relationships of QSs. Furthermore, we study the dynamical stability of these stars by analyzing the impact of anisotropy parameters $\beta$ and the interaction parameter $\lambda$ derived from the EoS, on their stability. Our results demonstrate that the presence of pressure anisotropy plays a significant role in increasing the maximum mass of QSs, with potential implications for the existence of super-massive pulsars. These findings are in agreement with recent astronomical observations, which suggest the possibility of neutron stars exceeding $2M_{\odot}$.
\end{abstract}

\maketitle

\section{Introduction}

The study of compact objects, such as QSs, has driven extensive research efforts due to their unique role as astrophysical laboratories for high-density matter and potential indicators of physics beyond general relativity (GR). Numerous modified gravity theories have been proposed to address fundamental cosmological questions, including dark energy and inflation. Among these theories, the Starobinsky model, which augments the Einstein-Hilbert action with a term proportional to the Ricci scalar squared, $f(R) = R + aR^2$, offers a simple yet compelling framework. This model not only addresses early-universe inflation but is also compatible with cosmic microwave background data, making it a promising candidate for investigating gravitational interactions at astrophysical scales \cite{Starobinsky:1980te, Akrami:2020bix}.

Astrophysical observations of supermassive neutron stars \cite{isChoi:2020eun,isChristian:2020xwz} and recent gravitational wave detections, such as GW190814 \cite{isLu:2020gfh}, suggest the need for theories that extend GR in strong-field regimes \cite{Abbott:2020khf}. These detections challenge traditional models, particularly in describing stellar structures under high densities and extreme conditions. In this context, the f(R) models \cite{Sotiriou:2008rp, DeFelice:2010aj, Nojiri:2010wj, Nojiri:2017ncd, Clifton:2011jh, Capozziello:2011et} and their particular realization in Starobinsky gravity offer insight into stellar stability and compactness properties that may deviate from GR predictions, thus making them attractive for modeling compact stars like QSs \cite{Astashenok:2020qds, Astashenok:2021peo, isTangphati:2024war, isTangphati:2024qkd, isTangphati:2024ycu, isTangphati:2024atj, Yazadjiev:2015zia, Astashenok:2017dpo, Blazquez-Salcedo:2018qyy}. The gravitational wave signal from GW 190814 has initiated the possibility on the secondary object as a neutron star with a mass 2.6 $M_{\odot}$. The modified gravity theories have been proposed on this phenomena \cite{Astashenok:2020qds, Astashenok:2021peo}

Another essential aspect of QS structure is pressure anisotropy, where the radial pressure, $P_r$, differs from the tangential pressure, $P_\perp$. Pressure anisotropy in compact stars is a natural consequence of various physical phenomena, including strong magnetic fields, rotation, and interactions among particles at ultra-high densities \cite{isRuderman:1972aj,isBowers:1974tgi}. In quark matter, pressure anisotropy affects the star's mass-radius relation, stability, and compactness, making it crucial to include such anisotropic features when studying QSs under modified gravity \cite{isTangphati:2024qkd,Herrera:1997plx}. Indeed, pressure anisotropy has been shown to significantly alter the maximum mass and radius of compact stars, highlighting its relevance in the structure of QSs modeled within the Starobinsky framework \cite{Banerjee:2020dad}.

This paper aims to investigate the effect of pressure anisotropy on the structural properties of QSs within the Starobinsky model. We employ a quark matter equation of state (EoS) tailored for anisotropic conditions, combined with the modified TOV \cite{isBauswein:2012ya} equations in Starobinsky gravity, to analyze the impact on mass-radius relationships and compactness. Our results extend existing studies on compact stars in f(R) theories by specifically addressing how anisotropy modifies the predictions of Starobinsky gravity regarding QS structure.

The plan of the paper is as follows: In Sec. \ref{sec2}, we formulate the problem and derive the field equations for Starobinsky's model under spherical symmetry. Section \ref{sec3} presents the anisotropic quark matter EoS used in this study. In Sec. \ref{sec4}, we discuss our numerical results, focusing on the mass-radius relationship and stability conditions under varying anisotropic parameters. Section \ref{sec5} provides concluding remarks and future perspectives. Throughout the paper, we adopt the signature $(-, +, +, +)$, and set $c = 1$.

\section{Field Equations  and set up}\label{sec2}

We begin with the most-studied modified gravity theory, so-called $f(R)$ gravity, in which Lagrangian
is replaced by a general function of the Ricci scalar, and the action is given by
\begin{eqnarray}\label{A}
S= \frac{1}{16\pi G} \int d^4x \sqrt{-g} f(R) + S_{\rm
matter}(\psi_i, g_{\mu\nu}),
\end{eqnarray}
where $f(R)$  is a function only of the Ricci scalar and 
where $g$ denotes the determinant of the metric $g_{\mu\nu}$. The $S_{\rm matter}$ denotes the action  of the matter field depending on the metric tensor and the matter field $\psi_i$. In order to prevent pathological scenarios like tachyonic instabilities and ghosts, the 
viable $f(R)$ theories must adhere to the following requirements \cite{Sotiriou:2008rp,DeFelice:2010aj}
\begin{eqnarray}\label{ine}
\frac{d^2f}{dR^2}\ge 0,  \;\;\; \frac{df}{dR}>0,
\end{eqnarray}
respectively. For our purposes, we shall consider a specific form of Starobinsky's model, i.e. $f(R)=R+aR^2$,  where  $a$ is taken to be  $a \geq 0$ in agreement with the inequalities given in Eq. \eqref{ine} \cite{Arapoglu:2010rz}. Here, the free parameter $a$ has dimensions of $[mass]^{-2}$, and we may write it in the form $a = 1/M^2$, stating that the mass scale $M$ is now the theory's free parameter. Selecting $a = 0$ allows us to obtain the GR solution again.  

In general, solving the fourth order differential equations in $4D$ spacetime is problematic. Thereby,   we adopt a conformal transformation by introducing the new scalar field $\varphi$ and the new metric $\tilde{g}_{\mu \nu}$ \cite{Brax:2008hh,Woodard:2006nt,Staykov:2014mwa}, which is 
\begin{equation}
\tilde{g}_{\mu \nu} = p g_{\mu \nu} = A^{-2}(\varphi) g_{\mu \nu},
\end{equation}
where $A (\varphi = \exp(-\phi/\sqrt{3})$.  Using this transformation one can rewrite the action (\ref{A}) in the Einstein frame as 
\begin{eqnarray}\label{act}
&& S = \frac{1}{16 \pi G} \int d^4x \sqrt{-\tilde{g}} [\tilde{R}-2 \tilde{g}^{\mu \nu} \partial_\mu \varphi \partial_\nu \varphi-V(\varphi)] \nonumber \\
&&+ S_M[\psi_i, \tilde{g}_{\mu \nu} A(\varphi)^2],
\end{eqnarray}
where the scalar-field potential $V(\varphi)$ takes the form
\cite{Brax:2008hh,Woodard:2006nt}  
\begin{equation}
V(\varphi) = \frac{(p-1)^2}{4 a p^2} = \frac{(1-exp(-2 \varphi/\sqrt{3}))^2}{4 a}.
\end{equation}

Taking variation with respect to the metric $\tilde{g}_{\mu \nu}$ and the scalar field $\varphi$, one arrives at the following modified field equation in the Einstein frame 
\begin{eqnarray} \label{feqs1}
&& \tilde{G}_{\mu \nu} = 8 \pi G [ \tilde{T}_{\mu \nu} + T^{\varphi}_{\mu \nu} ], \\
&& \nabla_\mu \nabla^\mu{\varphi}-\frac{1}{4} V_{,\varphi} = -4 \pi G \alpha (\varphi)\tilde{T}, \label{feqs2}
\end{eqnarray}
where $T^{\varphi}_{\mu \nu}$ is the the energy-momentum tensor corresponding to the scalar field and $V_{,\varphi} \equiv \frac{dV(\phi)}{d\phi}$. Moreover,  the coupling constant $\alpha (\varphi)$  is related with \cite{Staykov:2014mwa,Yazadjiev:2014cza}
\begin{eqnarray}
\alpha (\varphi)=\frac{d\ln{A(\varphi)}}{d\varphi}= -\frac{1}{\sqrt{3}}.
\end{eqnarray}

Through the formula $\tilde{T}_{\mu \nu} = A(\phi)^2 T_{\mu \nu}$, the energy-momentum tensor of the Einstein frame $\tilde{T}_{\mu \nu}$ is connected to that of the Jordan frame $T_{\mu \nu}$. We consider the matter source to be an anisotropic fluid, which
is described by 
 \begin{eqnarray}
T^{\nu}_i = (\rho+ P_{\perp})u^{\nu} u_{i} + P_{\perp} g^\nu_i + (P_r-P_{\perp})\chi_i \chi^\nu, 
\end{eqnarray}
where $\rho$ is the energy density, $P_{\perp}$ and $P_r$ are the pressures perpendicular and parallel to the
spacelike vector $\chi_\nu$. Then, the only nonzero diagonal components in the energy-momentum tensor are as follows: $T^\nu_i = \left( -\rho, P_r, P_{\perp}, P_{\perp} \right) $.

For an anisotropic fluid, the energy density, pressure components, and 4-velocity in the two frames are interconnected through the equations
\cite{Staykov:2014mwa,Yazadjiev:2014cza} 
\begin{eqnarray}
\tilde{\epsilon} & = & A(\phi)^4 \rho \\
\tilde{P} & = & A(\phi)^4 P_r \\
\tilde{P}_{\perp} & = & A(\phi)^4 P_{\perp} 
\end{eqnarray}
where the tilde indicates the Einstein frame. 

Now we look at a static spherically symmetric metric that describe non-rotating solutions of these field equations inside a compact star of $f(R)$ gravity,
\begin{eqnarray} \label{izmet}
ds^2= - e^{2\Phi(r)}dt^2 + e^{2\Lambda(r)}dr^2 + r^2 d \Omega^2,
\end{eqnarray}
where $d \Omega^2 = d\theta^2 + \sin^2\theta d\vartheta^2 $ is the line element on the unit 2-sphere. Moreover, all the metric functions depend only on the radial coordinate $r$.

Under these assumptions, we reach the following
set non-zero components of the field equations ({\color{black}for a more detailed derivation of the following field equations, see Appendix~\ref{app:derivation}.}), 
\begin{widetext}
\begin{eqnarray}
&&\frac{1}{r^2}\frac{d}{dr}\left[r(1- e^{-2\Lambda})\right]= 8\pi G
A^4(\varphi) \rho + e^{-2\Lambda}\left(\frac{d\varphi}{dr}\right)^2
+ \frac{1}{2} V(\varphi), \label{eq:FieldEq1} \\
&&\frac{2}{r}e^{-2\Lambda} \frac{d\Phi}{dr} - \frac{1}{r^2}(1-
e^{-2\Lambda})= 8\pi G A^4(\varphi) P_r +
e^{-2\Lambda}\left(\frac{d\varphi}{dr}\right)^2 - \frac{1}{2}
V(\varphi),\label{eq:FieldEq2}\\
&&\frac{d^2\varphi}{dr^2} + \left(\frac{d\Phi}{dr} -
\frac{d\Lambda}{dr} + \frac{2}{r} \right)\frac{d\varphi}{dr}= 4\pi G
\alpha(\varphi)A^4(\varphi)(\rho-P_r -2 P_{\perp})e^{2\Lambda} + \frac{1}{4}
\frac{dV(\varphi)}{d\varphi} e^{2\Lambda}, \label{eq:FieldEq3}\\
&&\frac{dP_r}{dr}= - (\rho + P_r) \left(\frac{d\Phi}{dr} +
\alpha(\varphi)\frac{d\varphi}{dr} \right)+{2 \over r}\left(P_{\perp} - P_r\right). \label{eq:FieldEq4}
\end{eqnarray}
\end{widetext}


{\color{black}
At the stellar surface ($r=R_S$) where $P_r=0$, the anisotropy term 
\begin{equation}\label{eq:anisotropy_term}
\frac{2}{r}(P_{\perp}-P_r)
\end{equation}
vanishes because our quasi-local ansatz, $\Delta = P_{\perp}-P_r = \beta P_r \zeta$, ensures that $P_{\perp}=0$ when $P_r=0$. This guarantees that the standard TOV boundary condition is preserved at the surface, while anisotropy continues to influence the star's interior structure.
}
In order to provide a comprehensive description of the configuration that is being examined, we must add an EoS for the fluid to the aforementioned equations. Note that the absence of a scalar field allows one to recover the standard conservation
equation. 

To close the modified TOV equations (\ref{eq:FieldEq1})-(\ref{eq:FieldEq4}) numerically, one needs boundary conditions for solving the interior and the exterior
problem simultaneously. Thus, we need boundary conditions in the natural Einstein frame to maintain regularity at the origin,
\begin{eqnarray}
\rho(0) = \rho_{c},\;\;\; \Lambda(0)=0,\;\;\;
\varphi(0) = \varphi_c, \;\;\; \frac{d\varphi}{dr}(0)=0,\label{eq:BC1}
\end{eqnarray}
where $\rho_c$ and $\phi_c$ are the central value of energy density and the scalar field, respectively, while at infinity
\begin{eqnarray}\label{BCINF}
\lim_{r\to \infty}\Phi(r) = 0, \;\;\;\; \lim_{r\to \infty}\varphi
(r)=0. \label{eq:BC2}
\end{eqnarray}
The radius of the star can be identified using the condition
\begin{eqnarray}\label{NSR}
P_r(r_S)=0.
\end{eqnarray}
where the radial pressure at the surface of the star vanishes.  Referring to \cite{Staykov:2014mwa}, the condition $\frac{d\varphi}{dr}(0) = 0$ guarantees the regularity of the scalar field $\phi$, which consequently secures the regularity of $\Phi$ at the stellar interior $r = 0$. Moreover, $\Lambda(0)=0$ is necessary for the Einstein frame geometry at the center to be regular. The Einstein and Jordan frame metrics are interconnected through a nonsingular conformal factor,  which guarantees the regularity of the Jordan frame geometry at the star's center. On the other hand, we choose the boundary conditions in such a manner that they could fulfill the asymptotic flatness criterion at infinity, which requires $\lim_{r\to\infty} V(\varphi(r)) = 0$, and gives $\lim_{r\to\infty}\varphi(r) = 0$. The criteria (\ref{BCINF}) guarantee asymptotic flatness in both the Einstein and Jordan frames.

As the coordinate radius of the star is determined through (\ref{NSR}), while the physical radius of the star, as measured in the physical Jordan frame, is expressed by
\begin{eqnarray}
R_{S}= A[\varphi(r_S)] r_S.
\end{eqnarray}

Next, we will show the structure equation for a QCD-driven EoS and EoS with an anisotropic profile that acts as a matter source for QSs in the Starobinsky model.




\section{Equations of State for Quark matter and the
anisotropy ansatz } \label{sec3}

\subsection{EoS for radial pressure}
\label{IQM_EOS}
The latest astronomical findings of massive pulsar masses support the possibility of a quark matter 
or QCD matter (quantum chromodynamic) core.  Therefore, both the theoretical and experimental communities have focused their efforts on QCD matter in extreme environments, such as high temperatures and/or density at the core of compact objects, for decades. Interestingly, the perturbative quantum chromodynamics (pQCD) formalism becomes more reliable as the quark mass increases \cite{Kapusta:2011}.
 This means that this method could be useful for studying stars that are very densely packed together. 

 The current work focuses on investigating the properties of compact stars made up of interacting quark matter (IQM) EoS, taking into account the superconducting effect~\cite{Alford:2002kj} and quantum chromodynamics (pQCD) corrections. Based on the correction terms, we rewrite the thermodynamic potential $\Omega$ as \cite{Alford:2004pf,isZhang:2020jmb}
\begin{align}
  \Omega=&-\frac{\xi_4}{4\pi^2}\mu^4+\frac{\xi_4(1-a_4)}{4\pi^2}\mu^4- \frac{ \xi_{2a} \Delta^2-\xi_{2b} m_s^2}{\pi^2}  \mu^2  \nonumber  \\
  &-\frac{\mu_{e}^4}{12 \pi^2}+B_{\rm eff} ,
\label{omega_mu}
\end{align}
where $\mu$ and $\mu_e$ are the respective average quark and electron chemical potentials. The unpaired free quark gas contribution is represented by the first term.  The pQCD contribution from one-gluon exchange for gluon interaction to $O(\alpha_s^2)$ order is expressed by the second term with $(1-a_4)$. The quartic coefficient $a_4$  is a constant accounting for the strong interactions between quarks \cite{Alford:2004pf}.  In the above expression, the term with $m_s$ accounts for the correction from the finite mass of the strange quark, if relevant.  Moreover, the term with the gap parameter $\Delta$ represents the contribution from color superconductivity. Different states of color-superconducting phases are represented by $(\xi_4, \xi_{2a}, \xi_{2b})$. Additionally, the effective bag constant, $B_{\rm eff}$, accounts for the nonperturbative contribution from the QCD vacuum. Generally the range of $B_{\rm eff}$ lies between $57 \geq B_{\rm eff} \geq 92$ MeV/fm$^3$ \cite{Blaschke:2018mqw, Burgio:2018mcr}.

Finally, the analytic expression related to energy density and radial pressure is \cite{isZhang:2020jmb,isZhang:2021fla}
{\fontsize{9.7}{12}
\begin{align}
P_r=\frac{1}{3}(\rho-4B_{\rm eff})+ \frac{4\lambda^2}{9\pi^2}\left(-1+{\rm sgn}(\lambda)\sqrt{1+3\pi^2 \frac{(\rho-B_{\rm eff})}{\lambda^2}}\right),
\label{eos_tot}
\end{align}}
where the constant coefficients $\lambda$  described by
\begin{eqnarray}
\lambda=\frac{\xi_{2a} \Delta^2-\xi_{2b} m_s^2}{\sqrt{\xi_4 a_4}}. \label{lam}
\end{eqnarray}
Note that $\text{sgn}(\lambda)$ represents the sign of $\lambda$. The focus of this investigation is solely on the positive $\lambda$ space \cite{isZhang:2020jmb}.

As demonstrated in Ref.~\cite{isZhang:2020jmb}, the $B_{\rm eff}$ parameter can be avoided using the subsequent dimensionless rescaling:
\begin{align}
\bar{\rho}=\frac{\rho}{4B_{\rm eff}}, \quad \bar{P}_r=\frac{P_r}{4B_{\rm eff}}, 
\label{rescaling_prho}
\end{align}
and
\begin{align}
\bar{\lambda}=\frac{\lambda^2}{4B_{\rm eff}}= \frac{(\xi_{2a} \Delta^2-\xi_{2b} m_s^2)^2}{4B_{\rm eff}\xi_4 a_4},
\label{rescaling_lam}
\end{align}
so that the EoS~(\ref{eos_tot}) reduces to the dimensionless form
\begin{align}
\bar{P}_r=\frac{1}{3}(\bar{\rho}-1)+ \frac{4}{9\pi^2}\bar{\lambda} \left(-1+{\rm sgn}(\lambda)\sqrt{1+\frac{3\pi^2}{\bar{\lambda}} {(\bar{\rho}-\frac{1}{4})}}\right).
\label{eos_p}
\end{align}

{\color{black}
The parameter $\bar{\lambda}$ in the above rescaled EoS directly influences the stiffness of the EoS. Larger values of $\bar{\lambda}$ lead to a stiffer EoS by increasing the radial pressure at a given energy density. This increased stiffness enables the QSs to support more mass and, as to be seen in Fig. \ref{fig_vary_lambdaBar} and Table \ref{table_vary_lambda}, results in an increase in both the maximum mass (from about $2.301\,M_\odot$ to $2.545\,M_\odot$) and radius. Consequently, the compactness parameter $M/R$ is slightly enhanced (rising from 0.287 to 0.294), while still remaining within the Buchdahl limit. This behavior reflects how the interaction strength encoded in $\bar{\lambda}$ plays a crucial role in determining the macroscopic properties of QSs.
}

Notice that as $\bar{\lambda} \to 0$, the aforementioned expression represents the rescaled conventional noninteracting quark matter EoS, where $\bar{P}_r=\frac{1}{3}(\bar{\rho}-1)$. Moreover, sufficiently large value of $\bar{\lambda}$ i.e., when $\bar{\lambda}>>0$, we get
\begin{align}
\tilde{P}_r\vert_{\tilde{\lambda}\to \infty}= \tilde{\rho}-\frac{1}{2}, 
\label{eos_infty}
\end{align}
or, equivalently, $P_r={\rho}-2B_{\rm eff}$, using Eq.~(\ref{rescaling_prho}). In Refs. \cite{isZhang:2020jmb,isZhang:2021fla,Pretel:2023nlr}, authors have shown  that the EoS (\ref{rescaling_prho}) and a much wider range of $\lambda>0$ lead to more massive QSs that meet the 2$M_{\odot}$ restriction.  \\
\\
Here, we examine the anisotropic QSs and particularly explore how we can incorporate the anisotropy as an extra matter component 
in the modified TOV equations  in addition with the interacting quark matter (IQM) EoS.  Starting with the quasi-local EoS \cite{Horvat:2010xf}, that has a broad range of applications in both modified gravity theory and GR. The explicit form of the EoS,  
\begin{eqnarray}\label{anisotropy}
    \Delta &\equiv& P_{\perp} - P_r =  \beta P_r \zeta.
\end{eqnarray}
where $\beta$ denotes the dimensionless free parameter that measures anisotropy. {\color{black}
This Ansatz is motivated by the physical conditions expected in ultra-dense compact stars. In this expression, the factor $P_r$ ensures that anisotropy scales with the local pressure and vanishes at the surface (where $P_r=0$), while the compactness parameter $\zeta=2m(r)/r$ captures the enhanced gravitational effects toward the star's core. Positive values of $\beta$ provide additional outward pressure that can counteract gravitational collapse, thereby enhancing stability, whereas negative values reduce the tangential pressure, potentially affecting stability in alternative ways. Our subsequent stability analysis confirms that this form of anisotropy maintains both the adiabatic indices above the critical threshold of $4/3$ and preserves causality throughout the stellar configuration.}

{\color{black}
Although the theoretical range for $\beta$ can extend from $-2$ to $2$, we restrict our study to $\beta \in [-0.35, 0.35]$ based on both physical and numerical considerations \cite{Doneva:2012rd,Silva:2014fca, Yagi:2015hda,Pretel:2020xuo, Rahmansyah:2020gar,Rahmansyah:2021gzt, Folomeev:2018ioy,Herrera:2013fja}. Our investigations indicate that values with $|\beta|>0.35$ tend to push the solutions toward the limits of physical acceptability-positive values bring the configuration uncomfortably close to the Buchdahl limit while negative values may compromise the stability by reducing the adiabatic indices below the critical threshold. This moderate range of $\beta$ is also more consistent with current microphysical models and observational constraints, ensuring that our quark star models remain realistic.
}

Additionally, the compactness is denoted as $\zeta \equiv 2m(r)/r$, where $m(r)$ is the gravitational mass bounded by radius $r$. As $r\to 0$, the anisotropy of pressures approaches zero, i.e., $\Delta \to 0$, resulting in the recovery of the isotropic solution. In \cite{Horvat:2010xf} it was shown that two pressure components vanish at the surface of the star, i.e., $P_r \left( r \rightarrow R \right) = P_{\perp} \left( r \rightarrow R \right) = 0$. Note that $\beta = 0$ maintains the regularity condition at the stellar interior also.



\section{Numerical results}\label{sec4}

Finally, we are in a position to solve the TOV equations (\ref{eq:FieldEq1})-(\ref{eq:FieldEq4})
within the scenarios of interacting quark matter (IQM) EoS (\ref{eos_p}) admixed with quasi-local EoS (\ref{anisotropy}). We perform a series of numerical calculation with different combinations of three parameters $(a,~ \beta, ~  , \Bar{\lambda})$, 
and examine the impact of these parameters on the mass-radius $(M-R)$ relationships. The conventional $(M, R)$ relation has been reconstructed by introducing $(M, R) = (\bar{M}/\sqrt{4B_{\rm eff}}, \bar{R}/\sqrt{4B_{\rm eff}})$. In the following, we measure the mass of the star in solar mass $(M_{\odot})$, radius is measured in $km$,
the energy density and the pressures in $MeV/fm^3$, and the parameter $a$ in $\text{ km}^2$, whereas the dimensions of the scalar field and the factor of compactness are dimensionless.

\begin{figure}
    \centering
    \includegraphics[width = 9.0 cm]{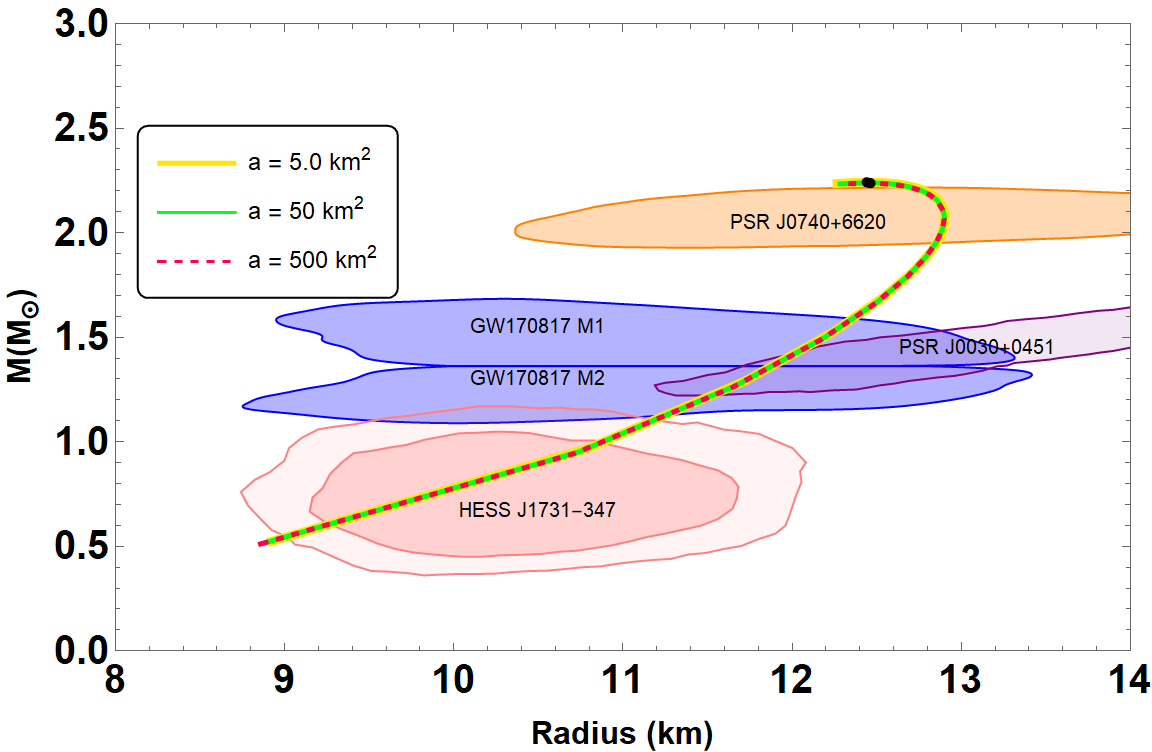}
    \includegraphics[width = 9.0 cm]{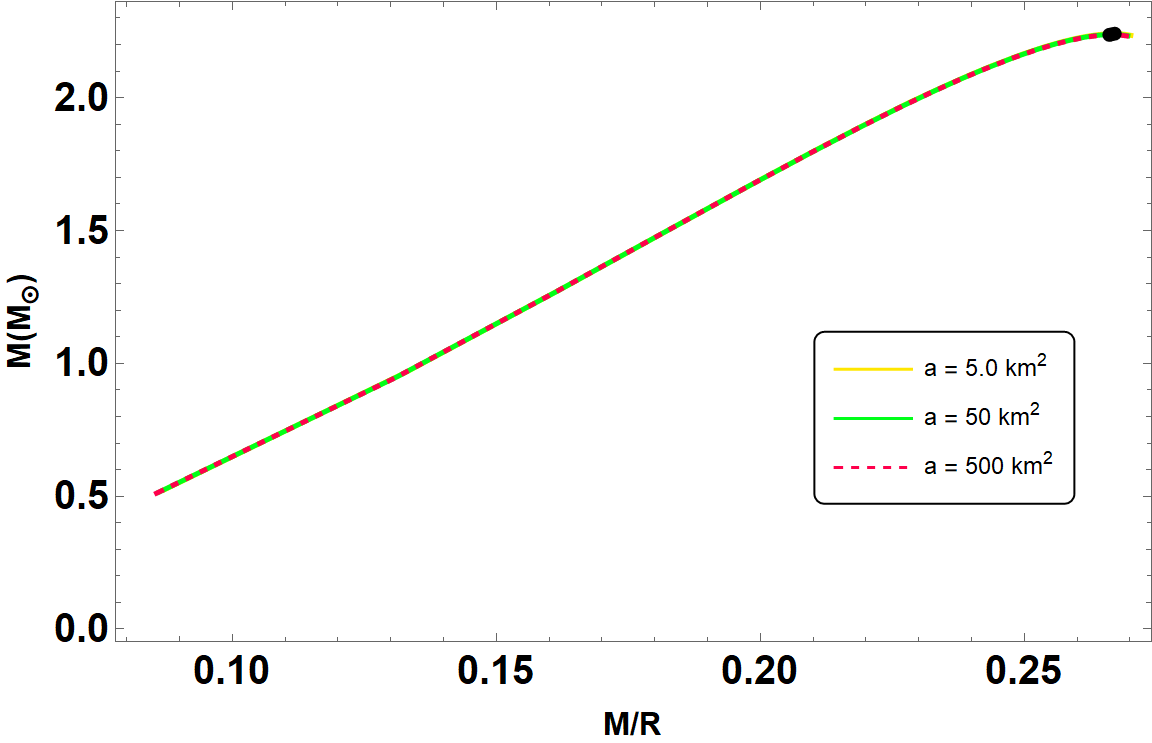}
    \caption{Mass-radius $(M-R)$ and mass-compactness $(M-M/R)$ relations are obtained within the scenarios of interacting quark matter (IQM) EoS (\ref{eos_p}) admixed with quasi-local EoS (\ref{anisotropy}). For the given central density, shown in Table \ref{tab:table1}, we plot all the diagrams for the given parameter set: $B = 60$ MeV/fm$^3$, $\Bar{\lambda} = 0.2$, $\beta = -0.35$ in variation of $a \in [5, 50, 500]$ km$^2$. \textcolor{black}{The shaded areas in the diagram represent observation constraints from analysis of NICER observations the pulsars PSR J0030+0451 \cite{Riley:2019yda}
 and PSR J0740+6620 \cite{Miller:2021qha}. }  The solid blue contour regions stand for the GW170817 event \cite{LIGOScientific:2018cki}, whereas
 the estimation for the mass and radius of the HESS J1731-347 \cite{Doroshenko:2022nwp} are shown in dark and light pink.}
    \label{fig_vary_a}
\end{figure}

\begin{table}[h]
\caption{\label{tab:table1} The fundamental features of QSs in $R^2$ gravity are presented. Parameters of the considered QSs are $B = 60$ MeV/fm$^3$, $\Bar{\lambda} = 0.2$, $\beta = -0.35$ in variation of $a \in [5, 50, 500]$ km$^2$.  }
\begin{ruledtabular}
\begin{tabular}{ccccc}
$a$    & $M_{max}$     & R    & $\rho_c$   & $M/R$\\
km$^2$ & $(M_{\odot})$ & (km) & MeV/fm$^3$ & \\
\hline
  5 & 2.247 & 12.470 & 923 & 0.267 \\[0.03cm]
  50 & 2.239 & 12.456 & 974 & 0.266 \\[0.03cm]
  500 & 2.237 & 12.463 & 974 & 0.266 \\[0.03cm]
\end{tabular}
\end{ruledtabular}
\end{table}



\begin{figure}
    \centering
    \includegraphics[width = 9.0 cm]{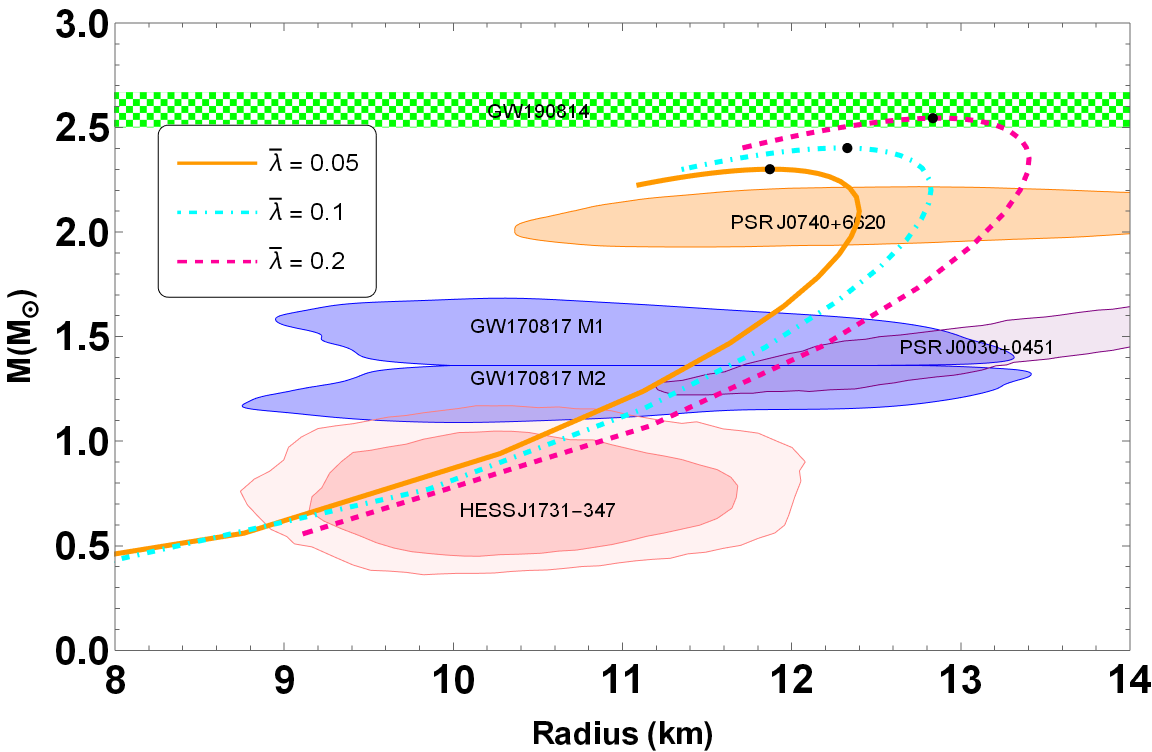}
    \includegraphics[width = 9.0 cm]{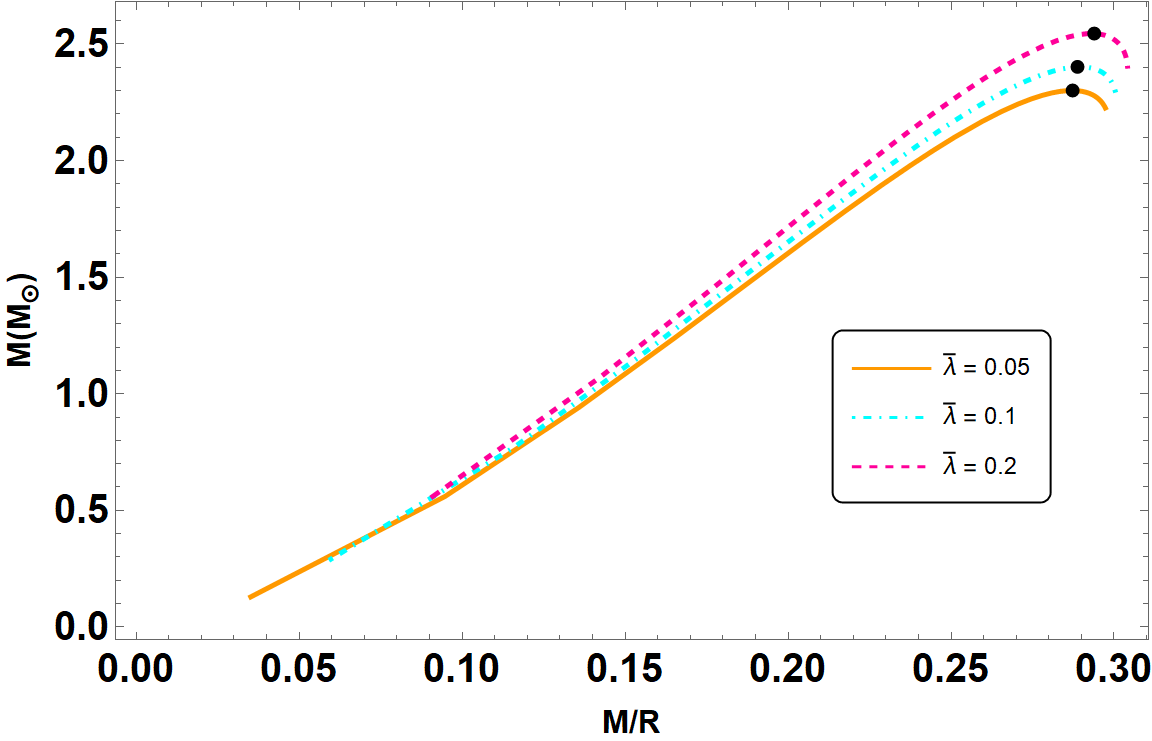}
    \caption{Mass-radius $(M-R)$ and mass-compactness $(M-M/R)$ relations are obtained within the scenarios of  interacting quark matter (IQM) EoS (\ref{eos_p}) admixed with quasi-local EoS (\ref{anisotropy}). For the given central density, shown in Table \ref{table_vary_lambda}, we plot all the diagrams for the given parameter set: $\beta = 0.35$, $B = 60$ MeV/fm$^3$, $a = 50.0$ km$^2$ in variation of $\Bar{\lambda} \in \{0.05, 0.10, 0.20 \}$. The presented observational constraints are same as of Fig. \ref{fig_vary_a}. \textcolor{black}{The $(M-R)$ diagram also includes mass measurements from the GW190425 event. \cite{LIGOScientific:2020zkf}.} }
    \label{fig_vary_lambdaBar}
\end{figure}

\begin{table}[h]
\caption{\label{table_vary_lambda} The fundamental features of QSs in $R^2$ gravity are presented. Parameters of the considered QSs are $B = 60$ MeV/fm$^3$, $\beta = 0.35$, $a = 50.0$ km$^2$ in variation of $\Bar{\lambda} \in \{0.05, 0.10, 0.20 \}$.}
\begin{ruledtabular}
\begin{tabular}{ccccc} 
$\Bar{\lambda}$ & $M_{max}$     & R    & $\rho_c$   & $M/R$\\
                & $(M_{\odot})$ & (km) & MeV/fm$^3$ & \\
\hline
  0.05 & 2.301 & 11.871 & 1,024 & 0.287 \\[0.03cm]
  0.10 & 2.402 & 12.329 & 923 & 0.289 \\[0.03cm]
  0.20 & 2.545 & 12.834 & 873 & 0.294 \\[0.03cm]
\end{tabular}
\end{ruledtabular}
\end{table}

\begin{figure}
    \centering
    \includegraphics[width = 9.0 cm]{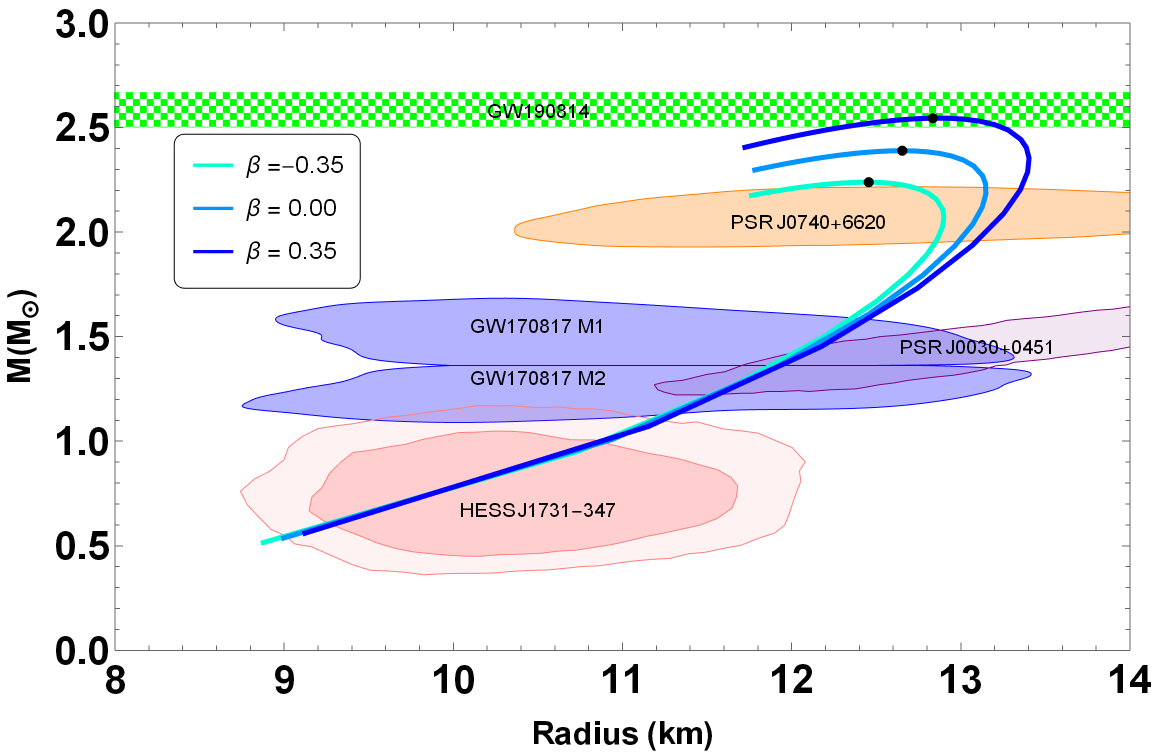}
    \includegraphics[width = 9.0 cm]{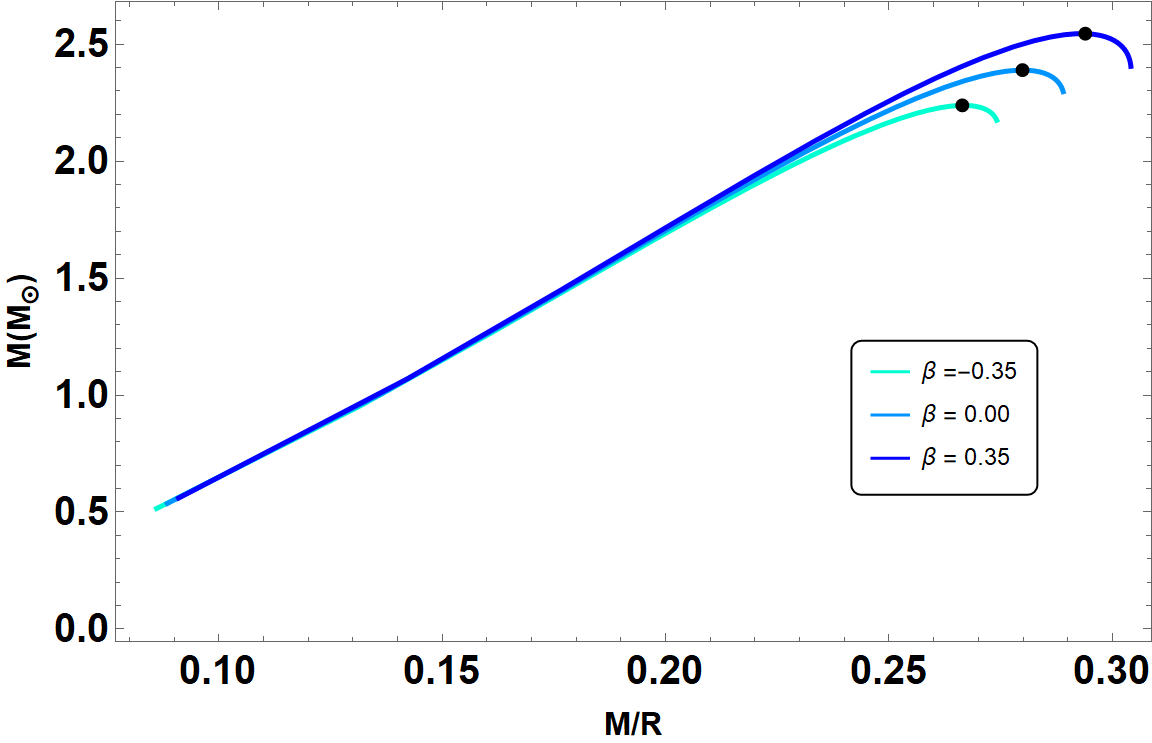}
    \caption{Mass-radius $(M-R)$ and mass-compactness $(M-M/R)$ relations are obtained within the scenarios of  interacting quark matter (IQM) EoS (\ref{eos_p}) admixed with quasi-local EoS (\ref{anisotropy}). For the given central density, shown in Table \ref{fig_vary_beta}, we plot all the diagrams for the given parameter set:  $B = 60$ MeV/fm$^3$, $\Bar{\lambda} = 0.2$, $a = 50.0$ km$^2$ in variation of $\beta \in \{-0.35, 0.00, 0.35\}$. The presented observational constraints are same as of Fig. \ref{fig_vary_a}. The green horizontal bar represents the maximum mass constraints from GW190814 event. }
    \label{fig_vary_beta}
\end{figure}

\begin{table}[h]
\caption{\label{table_vary_beta} The fundamental features of QSs in $R^2$ gravity are presented. Parameters of the considered QSs are $B = 60$ MeV/fm$^3$, $\Bar{\lambda} = 0.2$, $a = 50.0$ km$^2$ in variation of $\beta \in \{-0.35, 0.00, 0.35\}$. }
\begin{ruledtabular}
\begin{tabular}{ccccc}
$\beta$ & $M_{max}$     & R    & $\rho_c$   & $M/R$\\
        & $(M_{\odot})$ & (km) & MeV/fm$^3$ & \\
\hline
  -0.35 & 2.239 & 12.456 & 973 & 0.266 \\[0.03cm]
   0.00 & 2.389 & 12.653 & 923 & 0.280 \\[0.03cm]
   0.35 & 2.545 & 12.834 & 873 & 0.294 \\[0.03cm]
\end{tabular}
\end{ruledtabular}
\end{table}

\subsection{Profiles for Variation of $a$}\label{case1} 

To describe mass-radius $(M-R)$ and mass-compactness $(M-M/R)$ relations in Fig. \ref{fig_vary_a}, we vary the parameter $a \in [5.0, 50, 500]$ in km$^2$. To cover the parameter space of our model, we choose the other parameters as $B = 60$ MeV/fm$^3$, $\Bar{\lambda} = 0.2$ and $\beta = -0.35$, respectively. As one can see, the $(M-R)$ curves are almost indistinguishable from each other for different values of the parameter $a$ ranging from $a = 5$ to $a = 500$ km$^2$. This situation is similarly to Ref. \cite{Panotopoulos:2021sbf}, which supports the existence of anisotropic QSs with an integrating perturbative QCD corrections.  Table \ref{tab:table1} also includes a list of the properties of QSs, including their maximum masses and their corresponding radii.  From the table, it is seen that the maximum mass for QSs can be larger than 2$M_{\odot}$, and
goes up to 2.25 $M_\odot$ at $a = 5.0$ km$^2$ in our investigation. \textcolor{black}{ The orange and light purple areas correspond to the constraints on the mass and radius measurement derived  from 
the astrophysical observation data provided by the NICER  measurements of the pulsars PSR J0030+0451 \cite{Riley:2019yda} and PSR J0740+6620 \cite{Miller:2021qha}. } Moreover, the solid blue contour regions stand for the GW170817 event \cite{LIGOScientific:2018cki}, whereas
the estimation for the mass and radius of the HESS J1731-347 \cite{Doroshenko:2022nwp} are shown in dark and light pink. Beside that, we also demonstrate the impact of model parameter $a$ on the parameters of maximum compactness in the lower panel of Fig. \ref{fig_vary_a}. Notice that the trend of the $(M-M/R)$ curves is similar to that of $(M-R)$. This means that $(M-M/R)$ curves are almost indistinguishable from each other, and its value goes up to $0.266$, as evident from Table \ref{tab:table1}. {\color{black}
We note that the minimal differences observed in the mass-radius relation for varying values of $a$ (from 5 to 500 km$^2$) arise because the $R^2$ correction mainly influences the high-curvature core region. In particular, the corresponding scalar field decays rapidly outside this core, and the effective potential,
\begin{equation}
V(\varphi)=\frac{(1-\exp(-2\varphi/\sqrt{3}))^2}{4a},
\end{equation}
becomes flatter for larger $a$, thereby limiting its impact on the overall stellar structure. Moreover, the imposed boundary conditions further constrain the scalar field's influence, making the macroscopic properties, such as mass and radius, largely insensitive to $a$. This behavior appears to be a generic feature of $R^2$ gravity in the context of compact stars, suggesting that more sensitive observables (e.g., tidal deformability) may be needed to distinguish between different values of $a$.
}
Moreover, the table also indicates that the Buchdahl limit remains intact, i.e., $M/R < 4/9$ \cite{buchdahl}.

\subsection{Profiles for Variation of $\Bar{\lambda}$}\label{case2} 

 We further carry out the analysis in variation of $\Bar{\lambda}$ with other  parameters for our numerical calculations: $\beta = 0.35$, $B = 60$ MeV/fm$^3$ and $a = 50.0$ km$^2$, respectively. In Fig. \ref{fig_vary_lambdaBar}, we show the $(M-R)$ curves obtained for different values of $\Bar{\lambda}$
 and the main QS properties summarized in Table \ref{table_vary_lambda}.  It is evident that the tendencies of the maximum masses and their corresponding radii are increasing in conjunction with the increasing values of $\bar{\lambda}$. In this case, the maximum gravitational mass and their radius are found to be 2.30-2.55 and 11.87-12.83 km, respectively, for the considered parameter sets (refer Table. \ref{table_vary_lambda}), which is consistent with  GW190814 data. Moreover, in the lower panel of Fig. \ref{fig_vary_lambdaBar}, we also illustrate the impact of $\Bar{\lambda} $ on the properties of maximal compactness. Observe that the maximal compactness increases as $\Bar{\lambda}$ increases, typically falling within the range of 0.287 to 0.294, as also found in Table \ref{tab:table1}. The table further demonstrates that the Buchdahl limit is preserved, i.e.,  $M/R < 4/9$ \cite{buchdahl}.

\begin{figure}
    \centering
    \includegraphics[width = 9.0 cm]{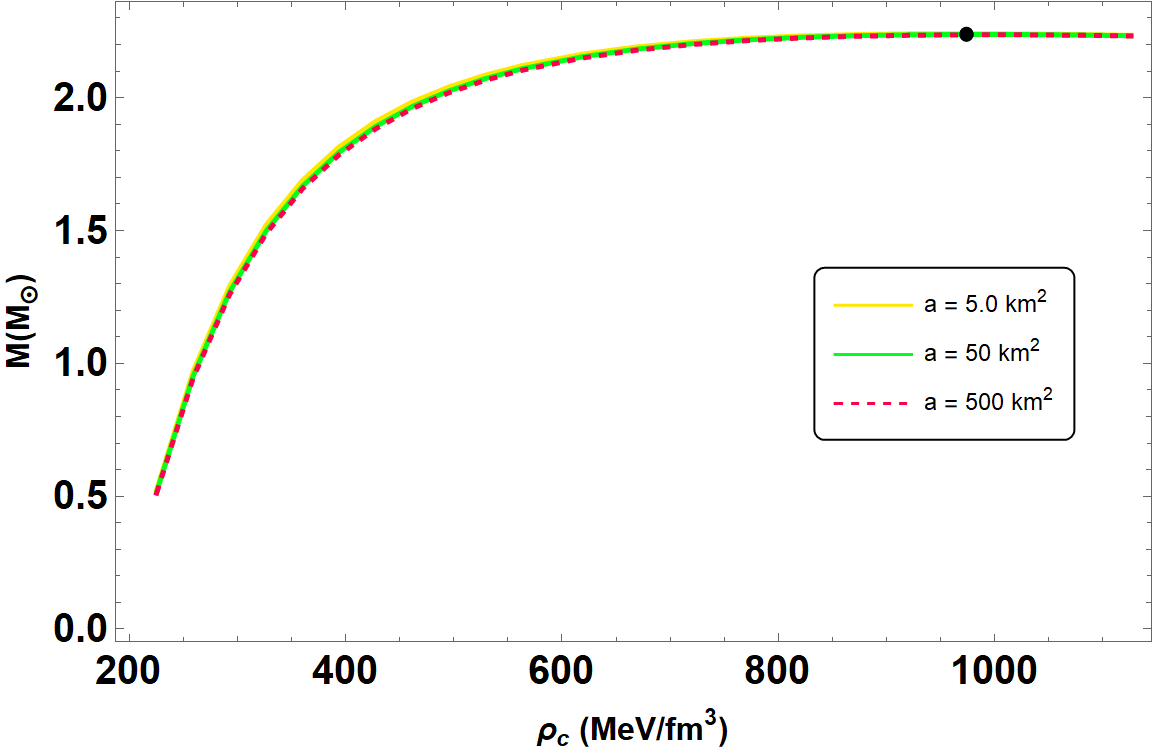}
    \includegraphics[width = 9.0 cm]{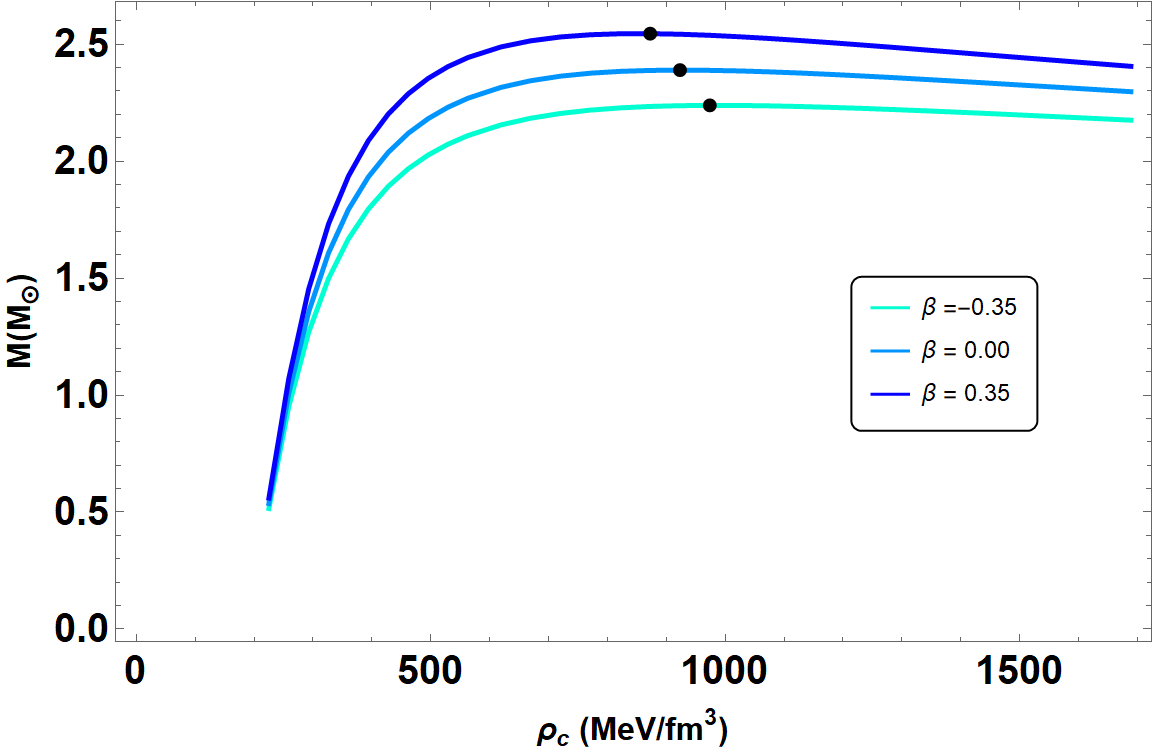}
    \includegraphics[width = 9.0 cm]{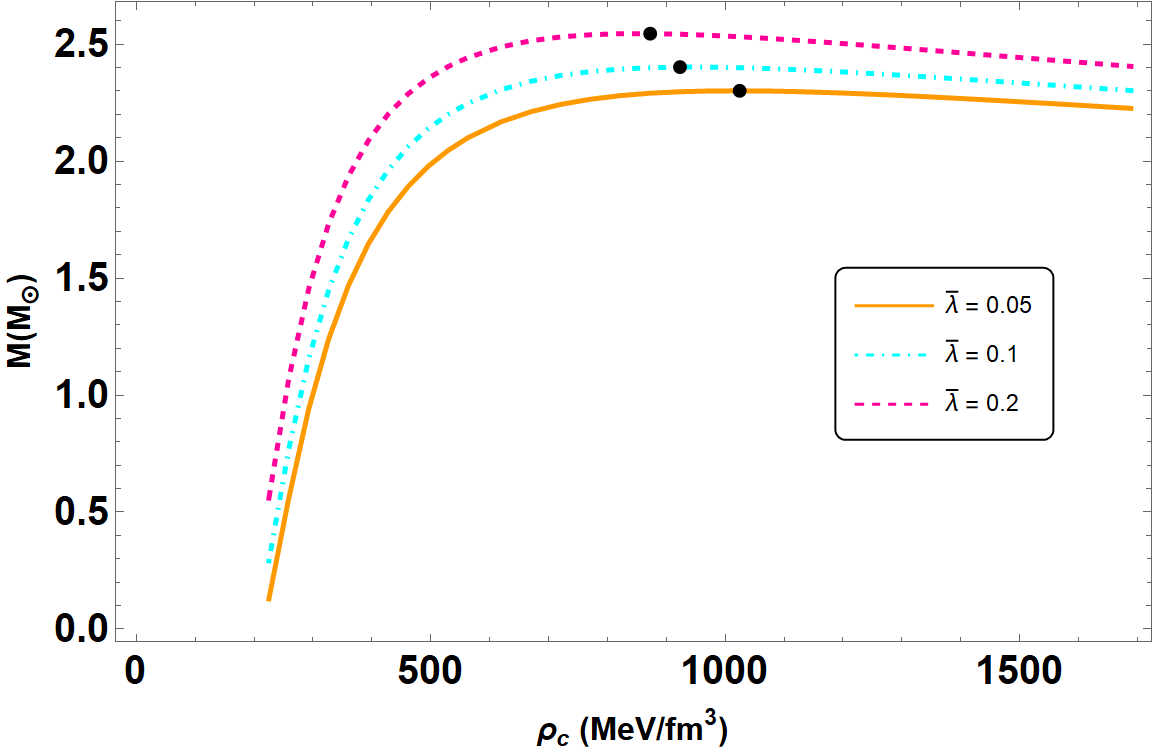}
    \caption{The profiles illustrate the relationships between mass and central density. The parameters used are the same as in Figs. \ref{fig_vary_a} to \ref{fig_vary_beta}.}
    \label{fig4}
\end{figure}

\begin{figure*}
    \centering
   \includegraphics[scale=.28]{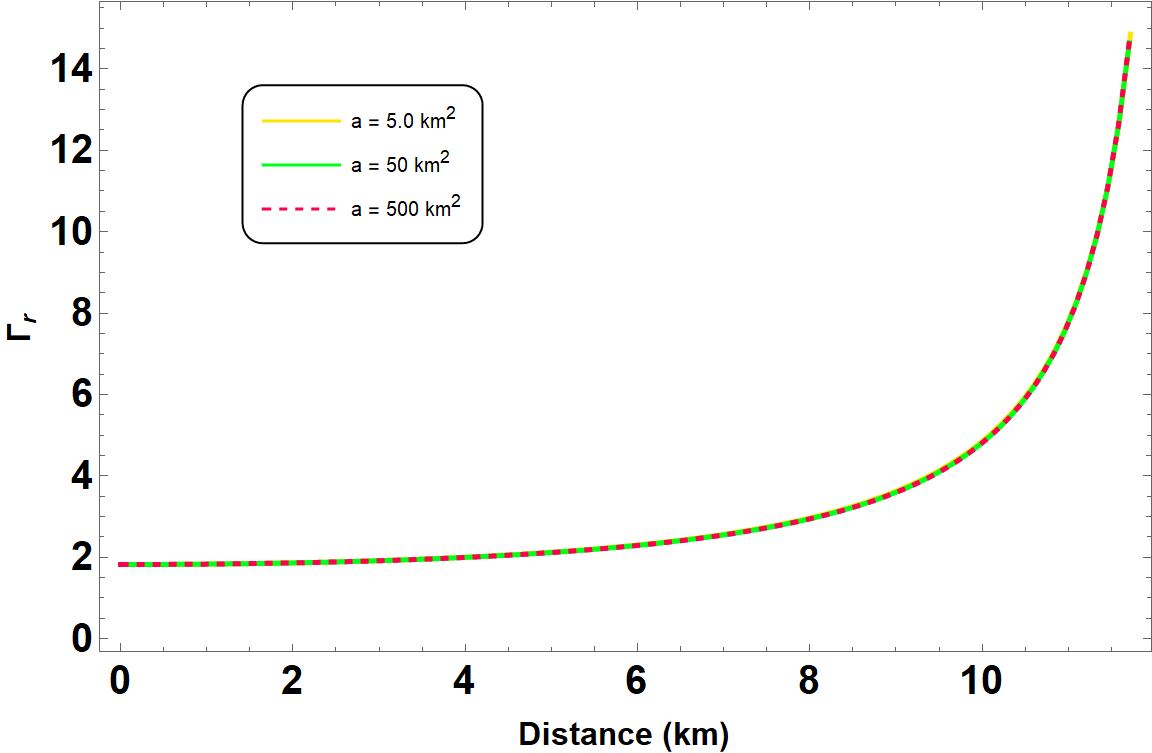}
    \includegraphics[scale=.28]{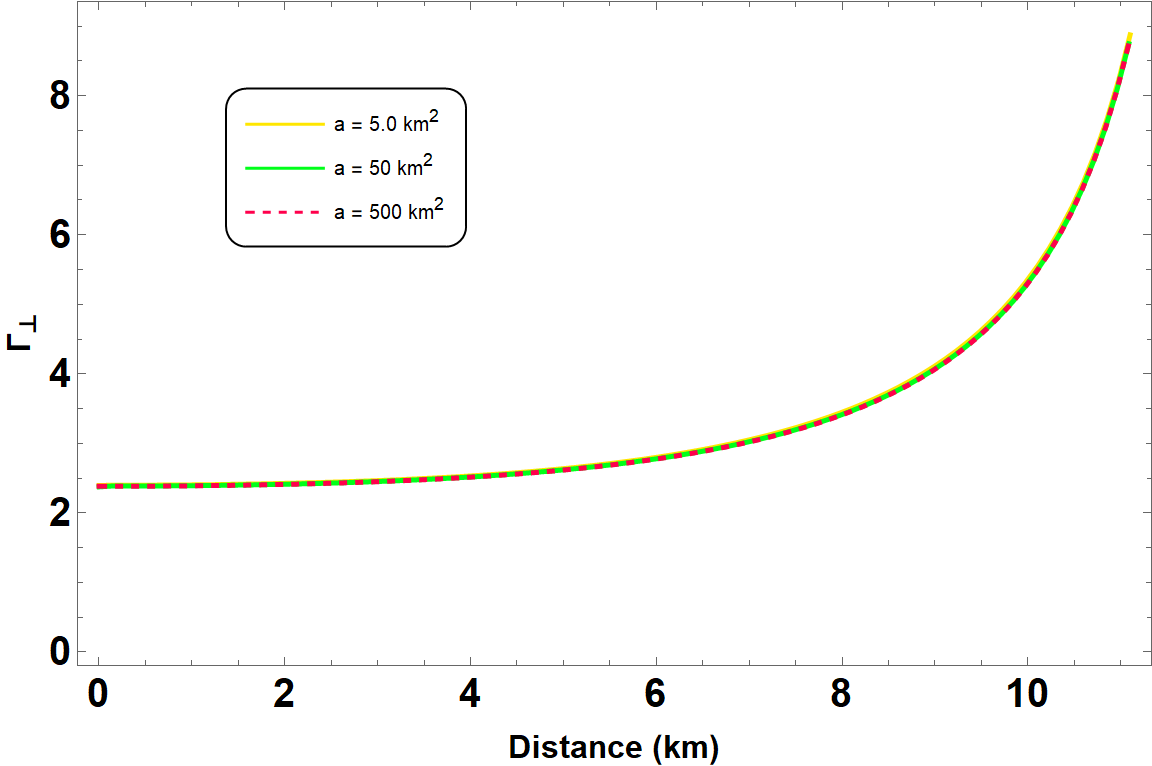}
    \includegraphics[scale=.29]{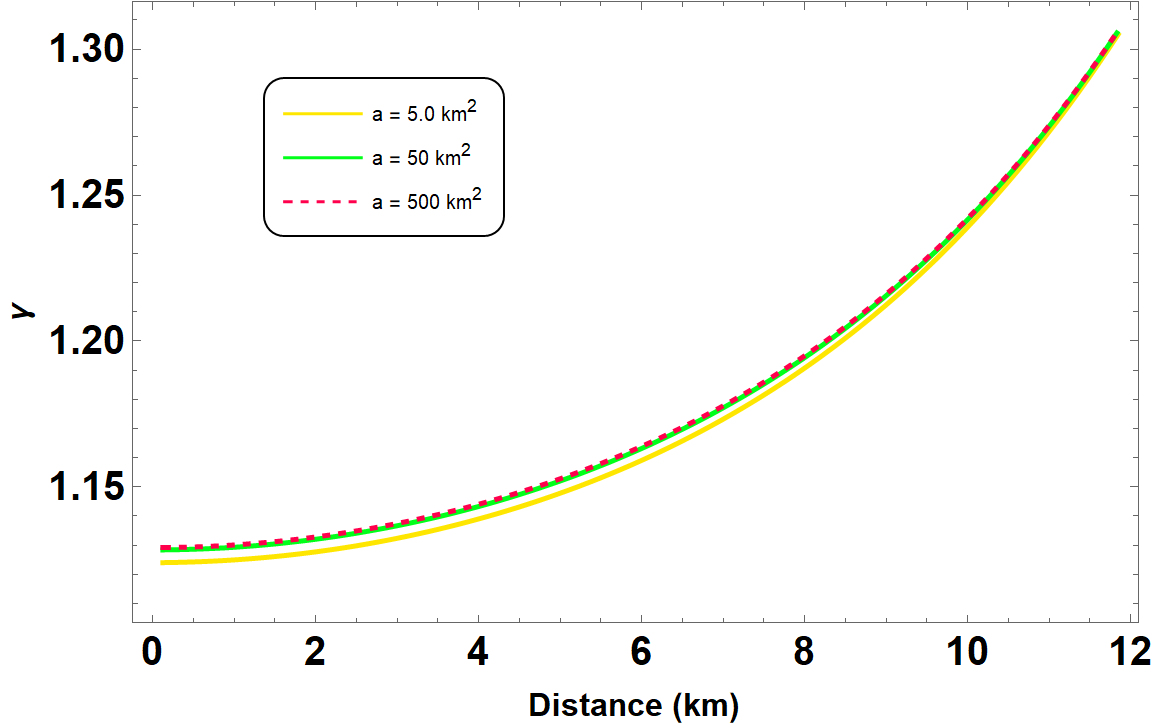}\\
    \includegraphics[scale=.28]{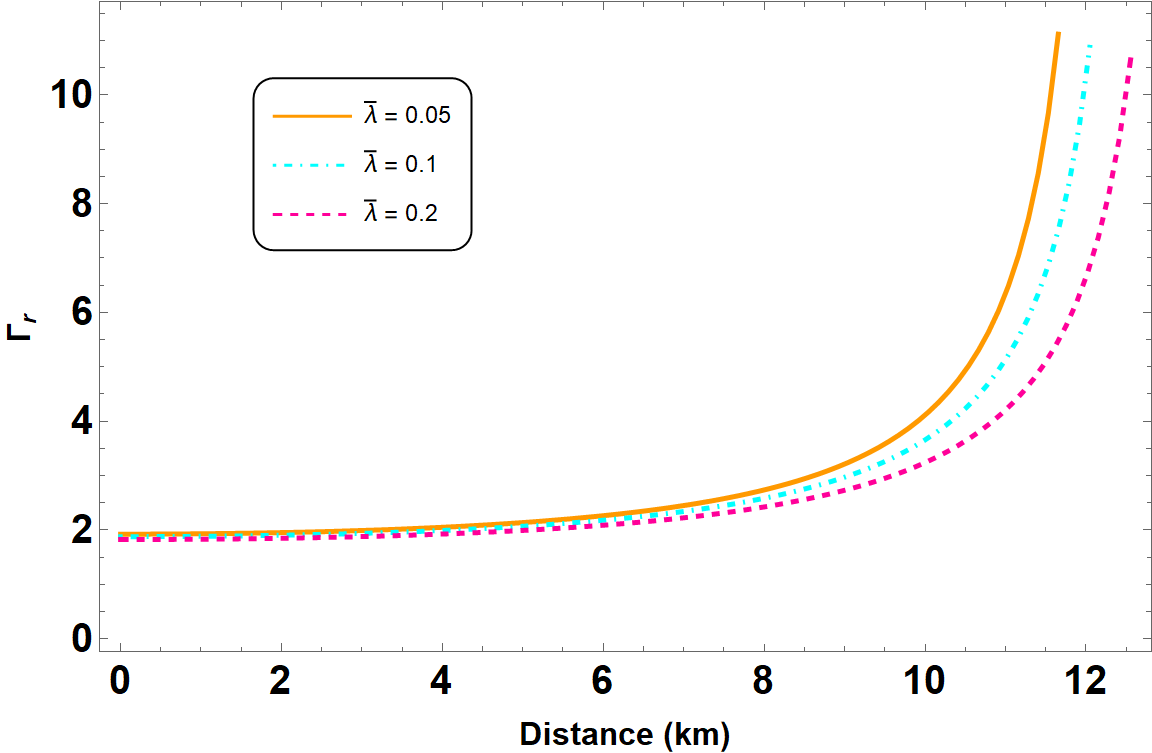}
    \includegraphics[scale=.28]{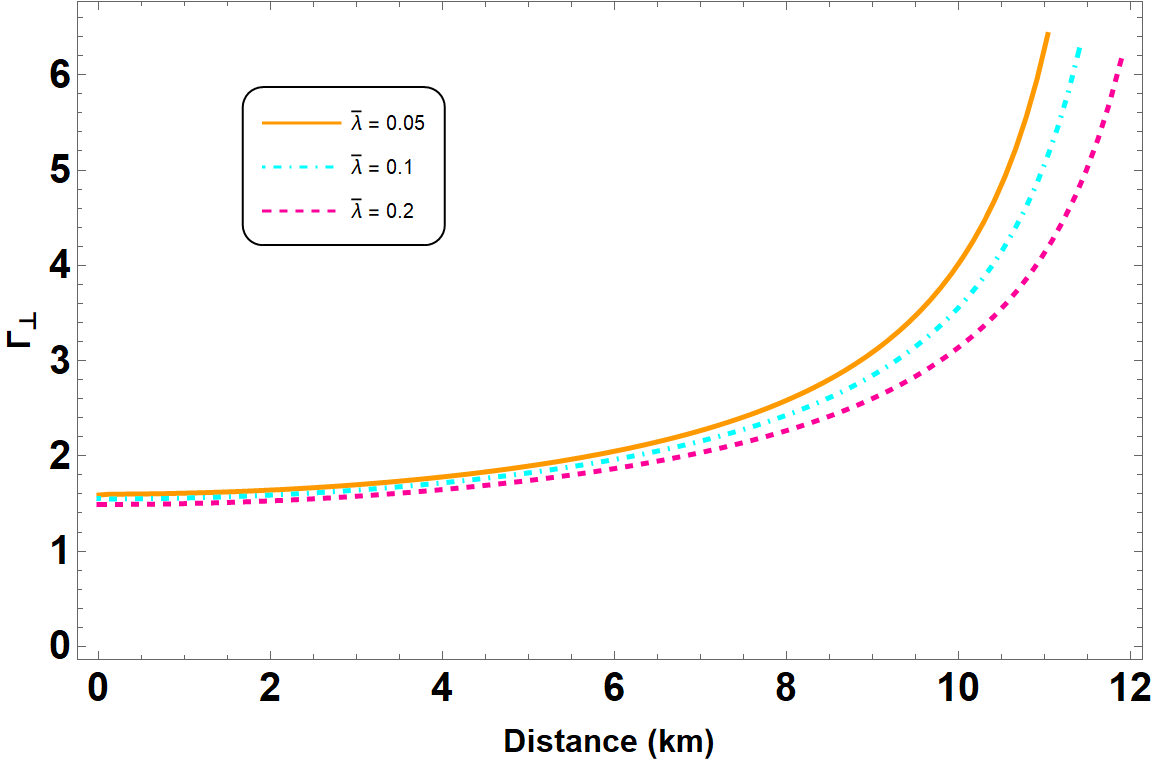}
    \includegraphics[scale=.29]{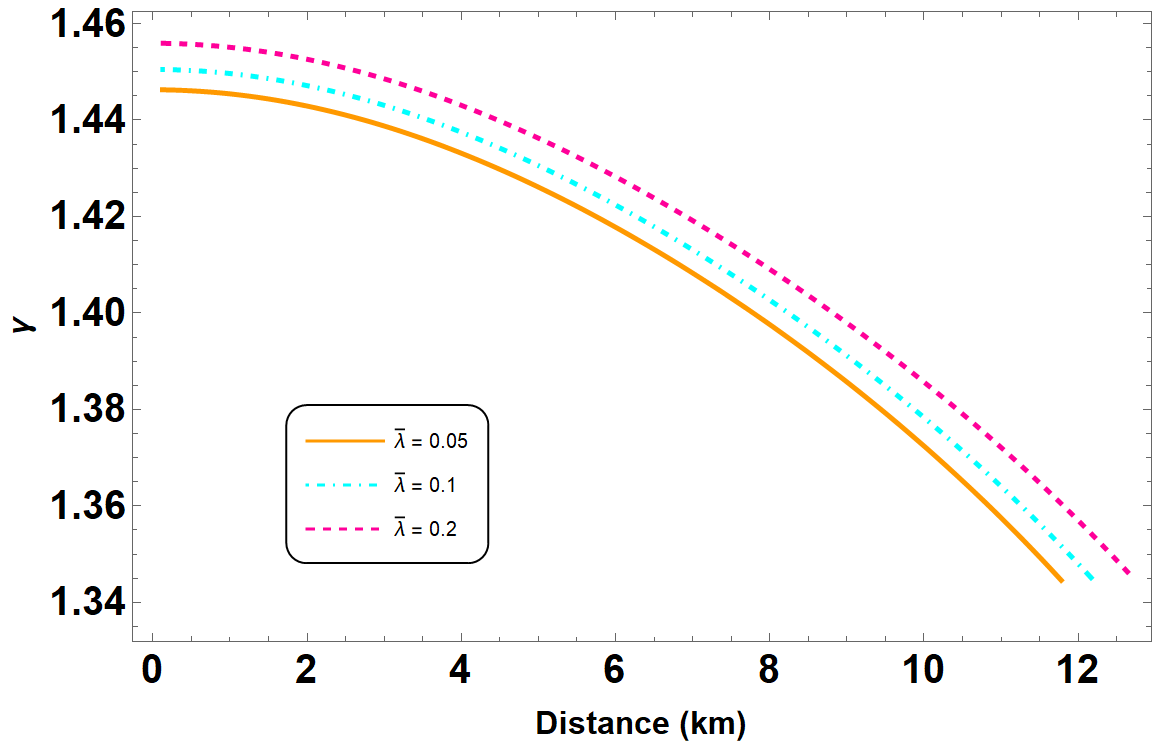}\\
    \includegraphics[scale=.28]{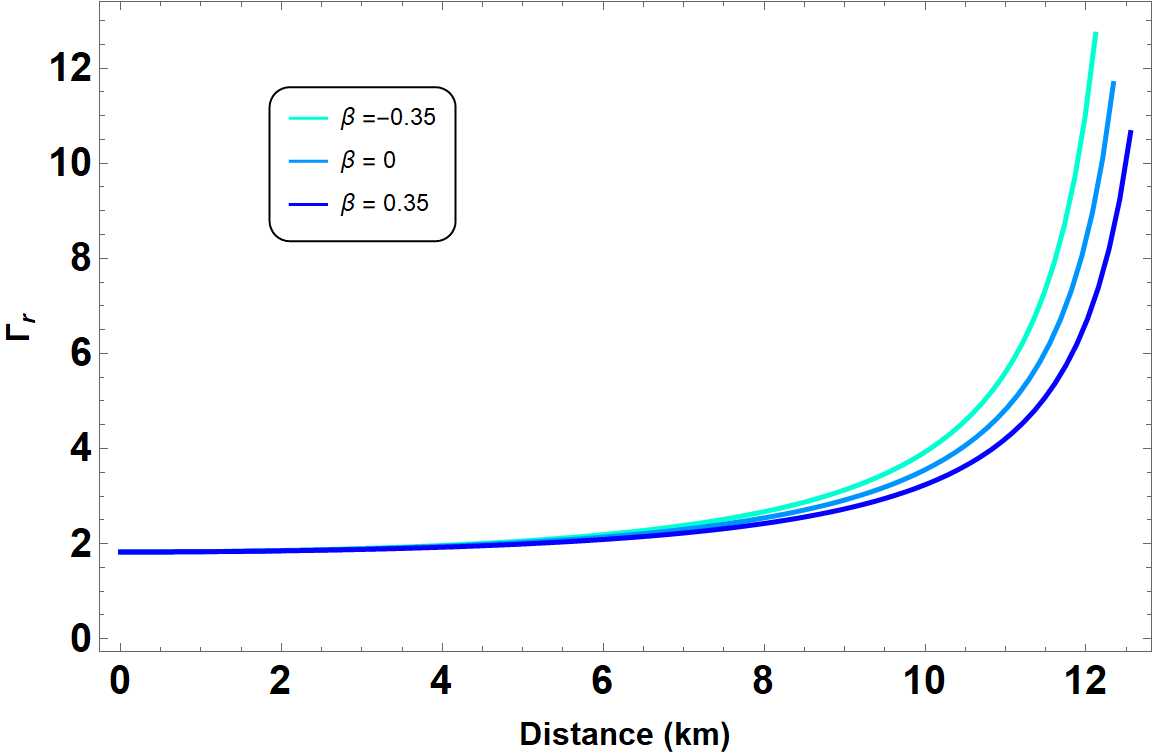}
   \includegraphics[scale=.28]{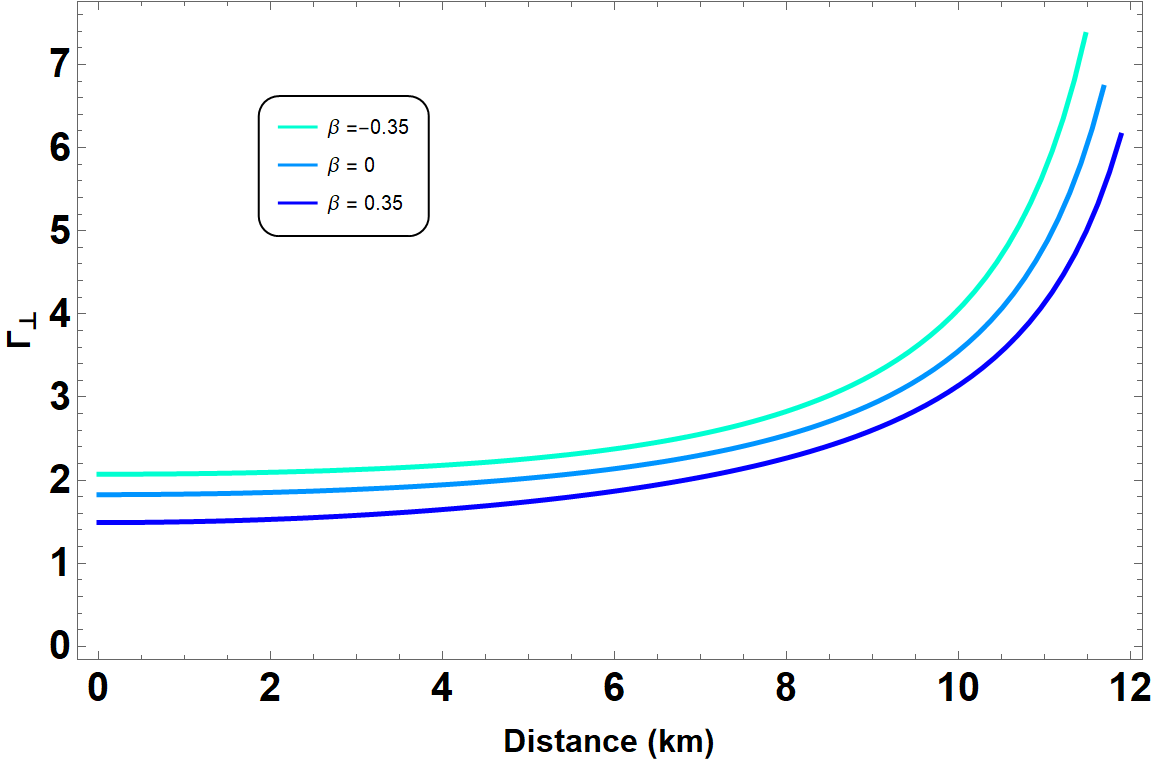}    \includegraphics[scale=.29]{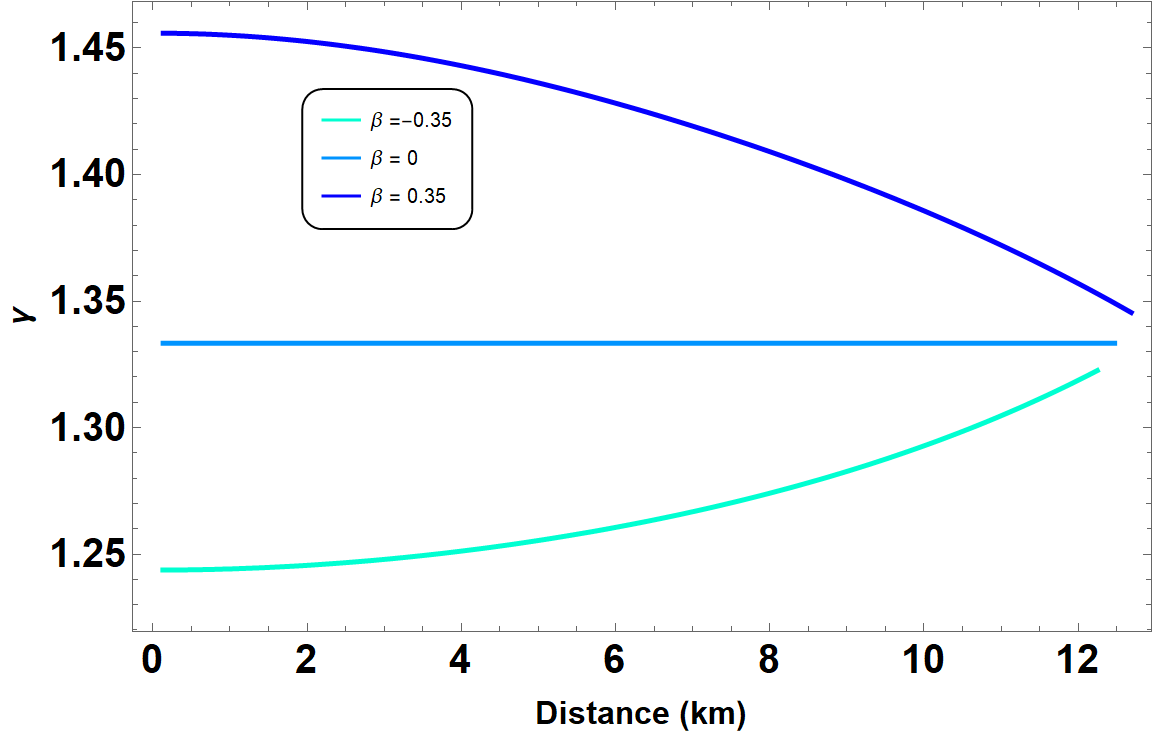}
    \caption{The adiabatic indices in radial ($\Gamma_r$) and tangential ($\Gamma_{\perp}$) directions with the same parameter set in \ref{fig_vary_a} to \ref{fig_vary_beta}.}
\label{fig_adia_a}
\end{figure*}

 \subsection{Profiles for Variation of $\beta$}\label{case3} 

Finally, we have examined the physical effect of anisotropy parameter $\beta$ on the properties of QSs. 
In the Fig. \ref{fig_vary_beta} we plot the $(M-R)$ and $(M-M/R)$ curves in variation of $\beta \in [-1, 1]$ 
with other  parameters for our numerical calculations: $B = 60$ MeV/fm$^3$, $\Bar{\lambda} = 0.2$ and $a = 50.0$ km$^2$, respectively. It is clearly seen from the figure that the $(M-R)$ curves vary significantly when the anisotropic pressure 
is varied, i.e., when we vary the parameter $\beta$. In principle, the maximum mass and radius of QSs more than
2$M_{\odot}$, and goes up to 2.55 $M_\odot$ at $\beta = 0.35$ from our calculation. This illustrates that there 
the maximum mass of the QS increases for increasing values $\beta$. Simultaneously, we also measured the QS's maximum mass 
for the isotropic scenario, or $\beta = 0$, which decreases to 2.34 $M_\odot$, see Table \ref{table_vary_beta}. This is a clear indication that the presence of the anisotropic pressure leads to the existence of a massive QS in $R^2$ gravity. Moreover, the theoretically attainable maximum mass comes out to be compatible with the mass-radius measurements obtained from different astrophysical observations. We further illustrate the $(M-M/R)$ diagram in the lower panel of Fig. \ref{fig_vary_beta}. Notice that the trend of $(M-M/R)$ curves follows the same pattern as $(M-R)$ relations; i.e., the maximum value of compactness increases when $\beta$ grows. This value lies within the range of  $0.266 < M/R < 0.294$. We summarize our results for various values of the parameter $\beta$ in Table \ref{table_vary_beta}. Interestingly,  the compactness does not exceed the Buchdahl bound, $M/R < 4/9$ \cite{buchdahl}.

\section{The static stability criterion, adiabatic index and the sound velocity}\label{sec5} 

We addressed the astrophysical feasibility of our proposed model, we will assess the stability of anisotropic QSs utilizing the static stability criterion, adiabatic index, and sound speed.  We shall examine each of them sequentially with a visual representation.

\subsection{Static stability criterion}

We will now examine the stability of the equilibrium configuration using the \textit{static stability criterion} (SSC) \cite{harrison,ZN}. It is a necessary condition, but not a sufficient one. The results are depicted in the $M-\rho_c$ plane, with $M$ representing the total mass and $\rho_c$ denoting the central density of the star. Mathematically, these inequalities are as follows: 
\begin{eqnarray}
\frac{d M}{d \rho_c} < 0 &~ \rightarrow \text{indicating an unstable configuration}, \\
\frac{d M}{d \rho_c} > 0 &~ \rightarrow \text{indicating a stable configuration}.
\label{criterion_M_rho_c}
\end{eqnarray}
In Fig. \ref{fig4}, we plot the variation of $M-\rho_c$ curves for all considered cases mentioned above. We utilize the same parameters for plotting as illustrated in Figs. \ref{fig_vary_a} and \ref{fig_vary_beta}. It is clear from those figures that the total gravitational mass
increases with growing central energy density, reaching a point where $(M_{\text{max}}, R_{M_{\text{max}}})$ is defined. Besides, the opposite inequality
$dM/d\rho_c <0$ implies instability with respect to small deformations. {\color{black} Moreover, Fig. \ref{fig4} shows that increasing the anisotropy parameter $\beta$ significantly modifies the stability boundary in the $M$-$\rho_c$ plane. In particular, the critical central density where $dM/d\rho_c=0$ shifts to lower values and the maximum stable mass increases (e.g., from $M_{\max}\approx2.24\,M_\odot$ at $\beta=-0.35$ to $M_{\max}\approx2.55\,M_\odot$ at $\beta=0.35$). This behavior arises because positive $\beta$ enhances tangential pressure, providing additional support against gravitational collapse, and leading to a steeper transition between stable and unstable configurations. These observations are consistent with our adiabatic index and sound speed analyses, confirming that the anisotropy plays a crucial role in stabilizing the QS configurations.
}

\subsection{Adiabatic Indices}

The adiabatic index $\gamma$ is a key parameter in determining the dynamical stability of compact stars, which was first proposed by Chandrasekhar \cite{Chandrasekhar}. For stability, the adiabatic index must satisfy the condition $\gamma > \frac{4}{3}$, which is critical for preventing gravitational collapse \cite{Glass}. The adiabatic indices in the radial ($\Gamma_r$) and tangential ($\Gamma_t$) directions are crucial in the case of anisotropic stars. They are defined as follows: \cite{Chan:gammaAdia, Nashed:2021sji, Nashed:2021gkp, Nashed:2022yfc, Nashed:2022zxm}

\begin{equation}
\gamma = \frac{4}{3} \left(1 + \frac{\sigma}{r |p_r'|} \right)_{\text{max}},
\end{equation}
where $\sigma = P_{\perp} - P_r$ is the anisotropy factor, and $p_r'$ is the derivative of the radial pressure. The stability condition depends on the anisotropy, as pressure behaves differently in radial and tangential directions. Therefore, we also define the following adiabatic indices for each direction:

\begin{equation}
\Gamma_r = \left(1 + \frac{\rho}{p_r}\right) v_r^2,
\end{equation}

\begin{equation}
\Gamma_t = \left(1 + \frac{\rho}{p_t}\right) v_t^2,
\end{equation}
where $v_r$ and $v_t$ are the sound speeds in the radial and tangential directions, respectively, $\rho$ is the energy density, $p_r$ is the radial pressure, and $p_t$ is the tangential pressure.

For isotropic stars, the adiabatic index $\gamma$ typically satisfies the stability criterion $\gamma > \frac{4}{3}$. However, in anisotropic stars, the condition $\gamma > \frac{4}{3}$ must be satisfied separately in both radial and tangential directions, which is why $\Gamma_r$ and $\Gamma_t$ become essential parameters in determining the stability.

According to Chandrasekhar's approach, the condition for stability is that both adiabatic indices $\Gamma_r$ and $\Gamma_t$ must exceed $\frac{4}{3}$ to ensure that the star is dynamically stable in both directions. In particular, anisotropy can enhance stability by increasing $\Gamma_t$, thereby providing additional resistance to collapse in the tangential direction.

Our results for the adiabatic indices in radial and tangential directions are shown in Figure \ref{fig_adia_a}, which indicate that the considered model is dynamically stable.  {\color{black}
Our analysis indicates that anisotropic effects on dynamical stability are parameter dependent. Moderate positive anisotropy (e.g., $0<\beta<0.35$) generally enhances stability by keeping both $\Gamma_r$ and $\Gamma_t$ safely above the critical value of $4/3$, whereas strong negative anisotropy (e.g., $\beta < -0.35$) may lead to instability in certain regions, particularly in the outer layers of the star.
}.


\begin{figure}
    \centering
    \includegraphics[width = 9.0 cm]{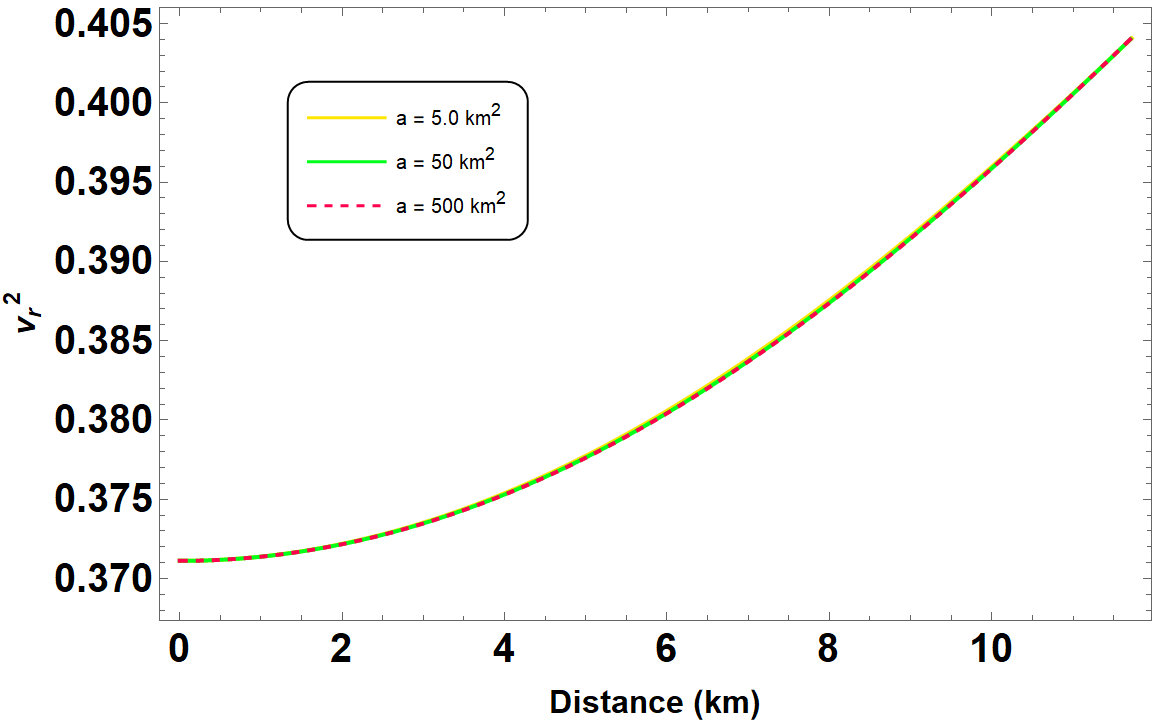}
    \includegraphics[width = 9.0 cm]{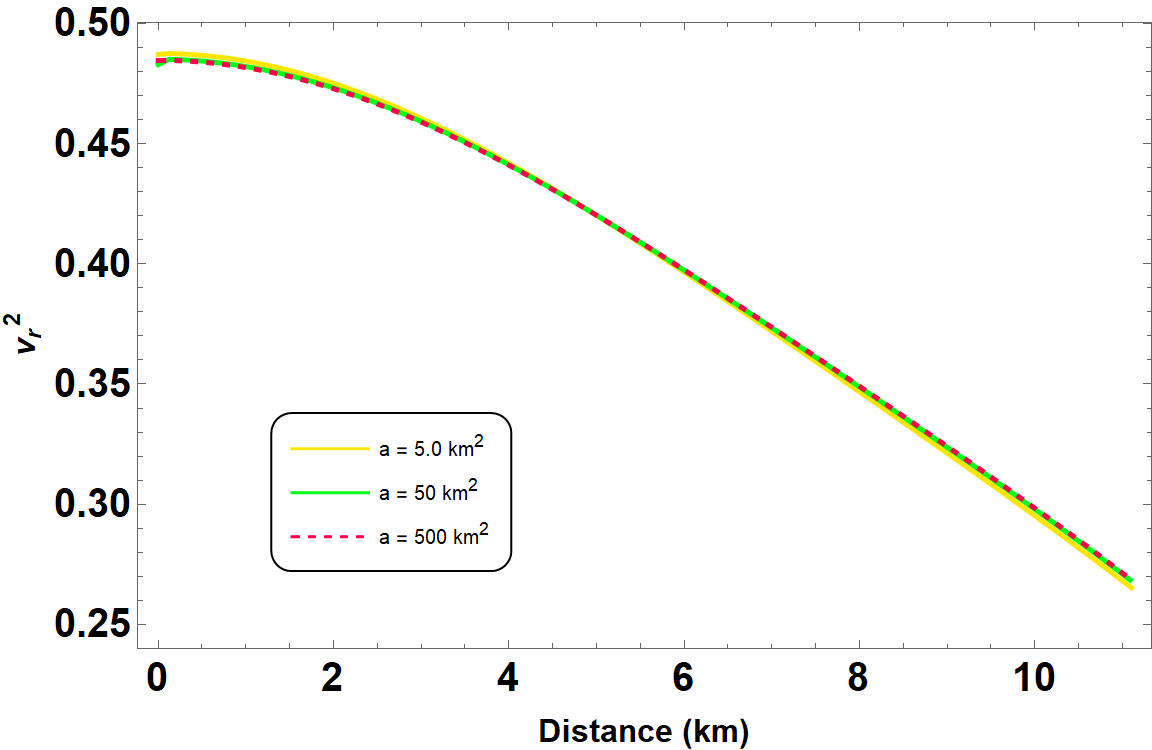}
    \caption{The square of sound speed has been plotted in both radial and tangential directions for the parameter sets defined in Fig. \ref{fig_vary_a}.}
    \label{fig_vel_a}
\end{figure}

\begin{figure}
    \centering
    \includegraphics[width = 9.0 cm]{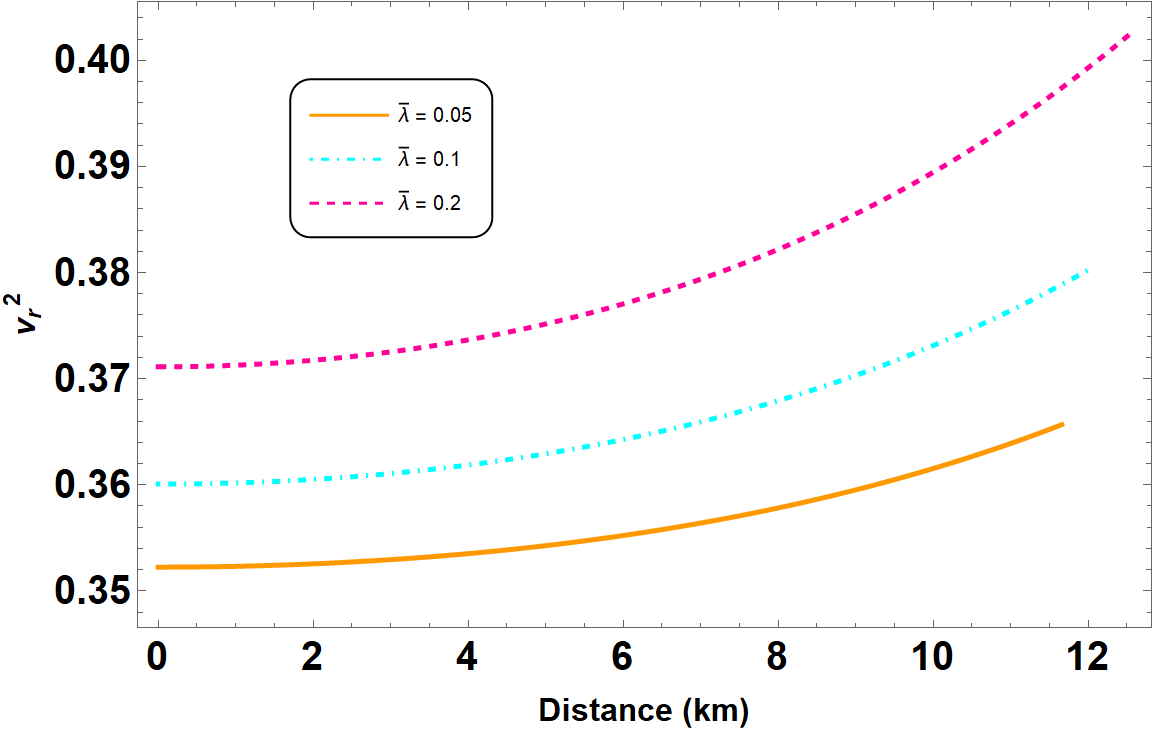}
    \includegraphics[width = 9.0 cm]{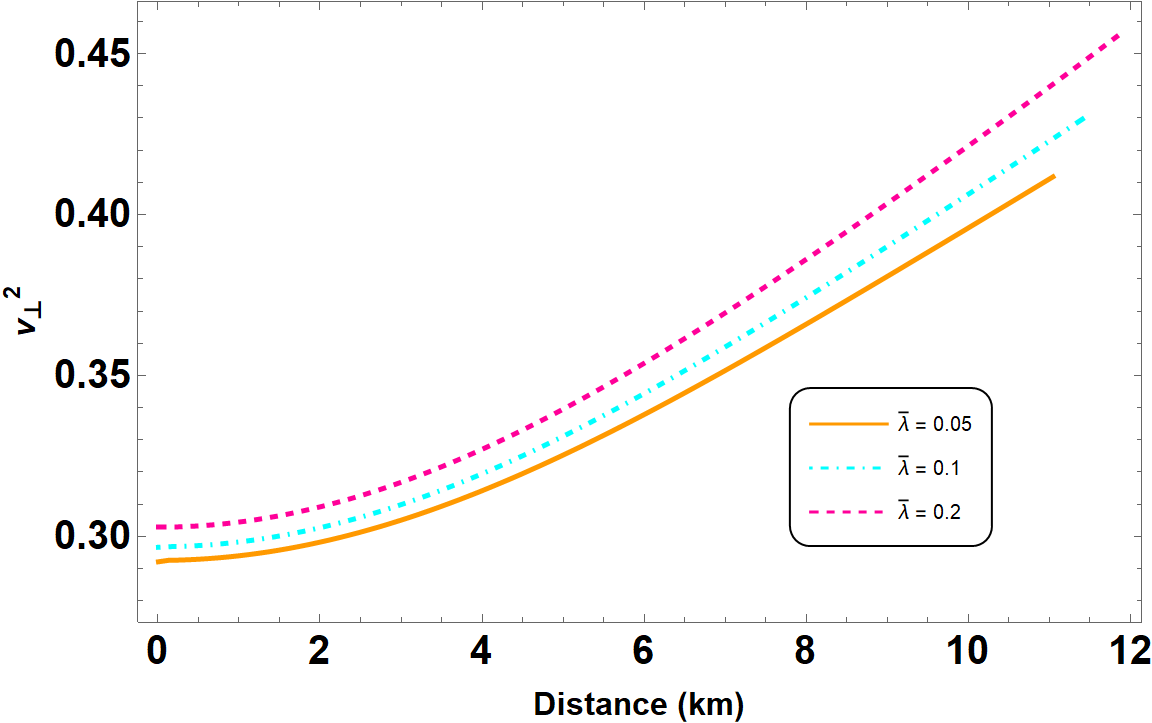}
    \caption{The square of sound speed has been plotted in both radial and tangential directions for the parameter sets defined in Fig. \ref{fig_vary_lambdaBar}.}
    \label{fig_vel_lambda}
\end{figure}

\begin{figure}
    \centering
    \includegraphics[width = 9.0 cm]{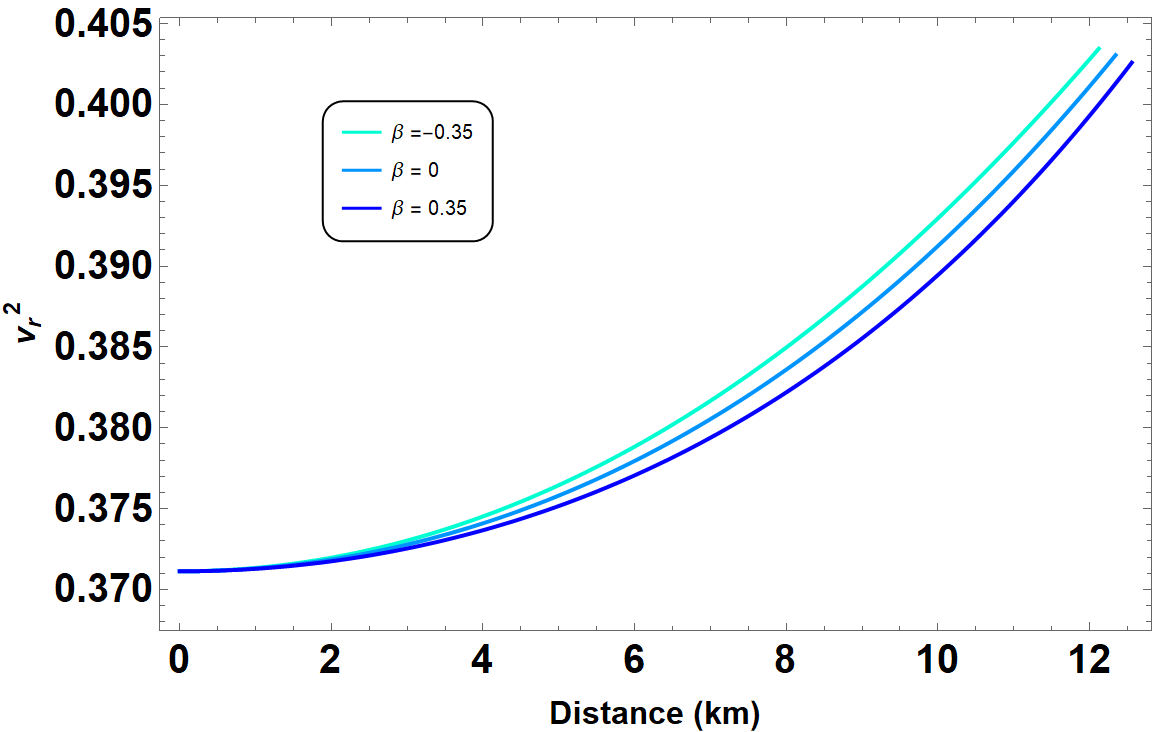}
    \includegraphics[width = 9.0 cm]{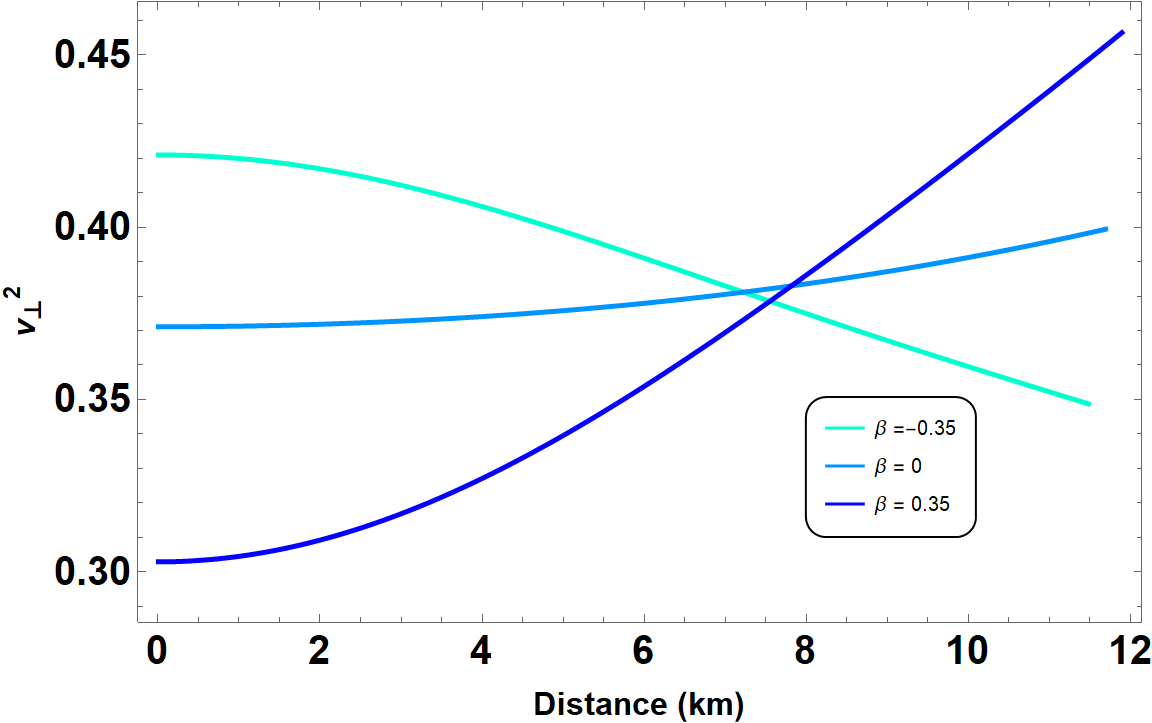}
    \caption{The square of sound speed has been plotted in both radial and tangential directions for the parameter sets defined in Fig. \ref{fig_vary_beta}.}
    \label{fig_vel_beta}
\end{figure}

\subsection{Sound speed and causality}

The sound speed $v_s$ is an important criterion for assessing both the stability and physical validity of stellar configurations. In general relativity, the sound speed must respect causality, meaning it must not exceed the speed of light ($c$). This imposes the condition:
\begin{equation}
0 \leq v_s^2 = \frac{dP}{d\rho} \leq 1,
\end{equation}
where $v_s^2$ is the square of the sound speed, $P$ is the pressure, and $\rho$ is the energy density. The inequality ensures that perturbations in the star propagate at subluminal speeds, preserving causality, and that the pressure is a monotonically increasing function of the energy density.

In our analysis, we compute the sound speed in both the radial ($v_r^2$) and tangential ($v_t^2$) directions to account for possible anisotropy in the stellar model. For the radial direction, the sound speed is defined as:
\begin{equation}
v_r^2 = \frac{dP_r}{d\rho}.
\end{equation}

Similarly, the tangential sound speed is given by:
\begin{equation}
v_{\perp}^2 = \frac{dP_{\perp}}{d\rho}.
\end{equation}

In the case of anisotropic stars, both $v_r^2$ and $v_{\perp}^2$ must independently satisfy the causality condition.

Figures \ref{fig_vel_a}, \ref{fig_vel_lambda}, and \ref{fig_vel_beta} illustrate the radial and tangential sound speeds for different parameter sets. Our results confirm that the sound speeds in both directions remain within the physically acceptable range, $0 \leq v_s^2 \leq 1$, thus satisfying the causality condition across all models considered. Additionally, we find that the variation in sound speed is smooth, indicating that the stellar configurations are stable and consistent with the physical requirements for compact stars.

Therefore, the computed sound speeds validate the stability and causal nature of our anisotropic QS models within the framework of $R^2$ gravity. These results also support the conclusion that the presence of anisotropy, characterized by different pressures in the radial and tangential directions, does not lead to violations of causality or instabilities in the QS configurations considered.

\section{Tidal deformation}\label{tidal}

\begin{figure}
    \centering
    \includegraphics[width = 9 cm]{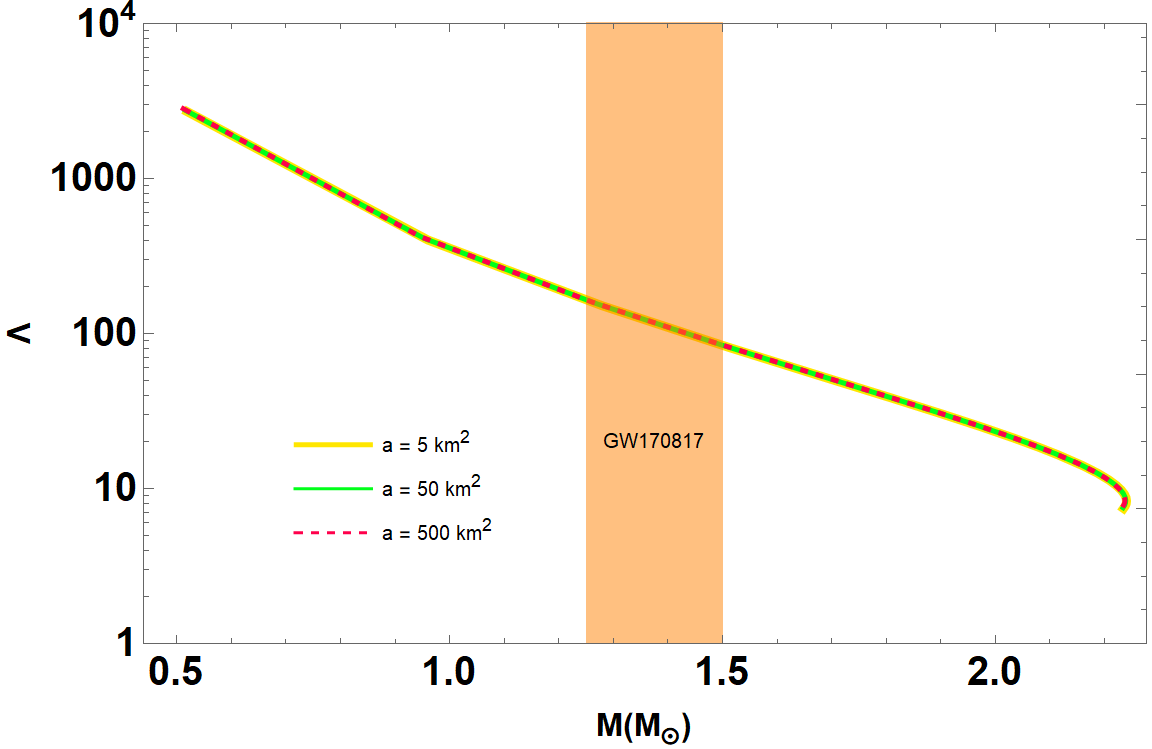}
    \includegraphics[width = 9 cm]{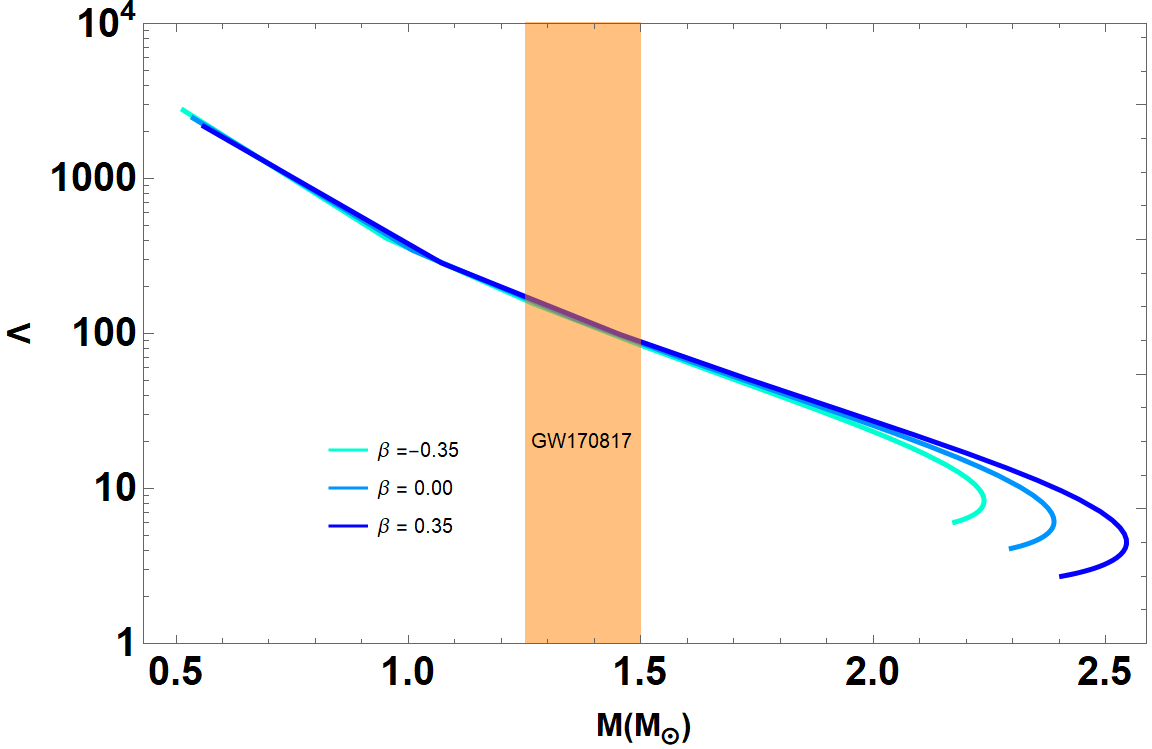}
    \includegraphics[width = 9 cm]{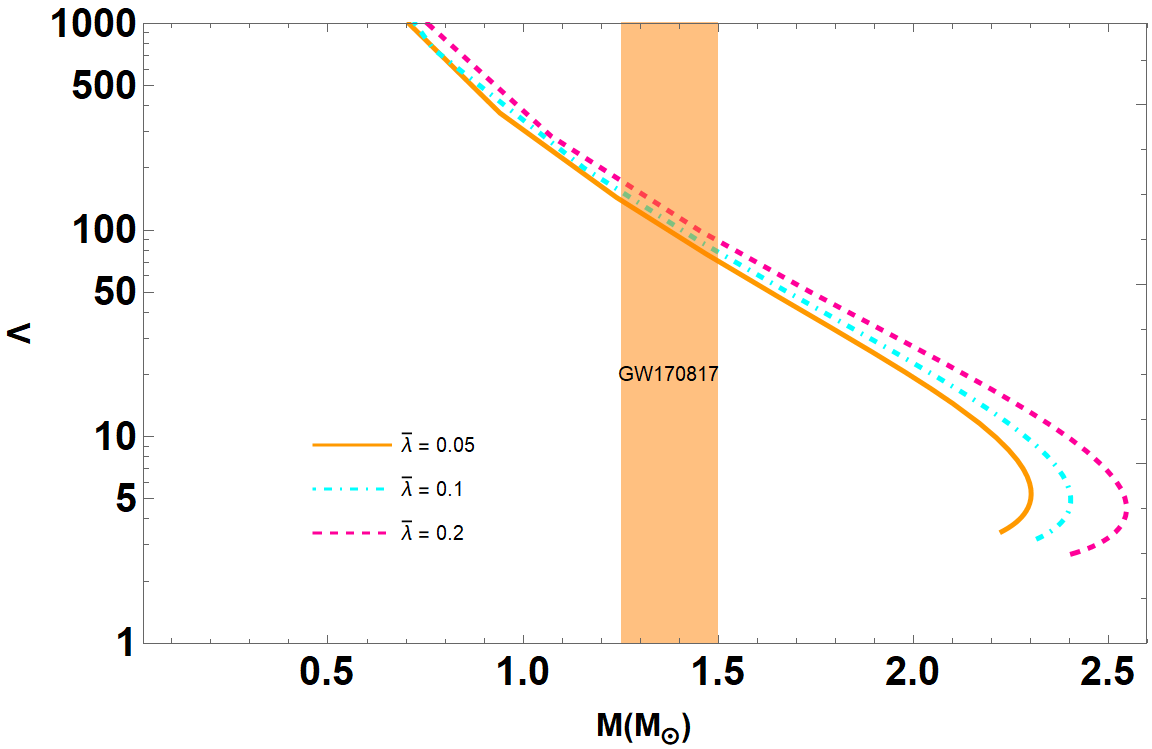}
    \caption{The tidal deformation of the anisotropic QSs in Starobinsky model has been plotted for the parameter sets defined in Fig.~\ref{fig_vary_a}, \ref{fig_vary_lambdaBar} and \ref{fig_vary_beta}, respectively.}
    \label{fig_tidal}
\end{figure}

\textcolor{black}{
Due to the study of gravitational waves, physicists can investigate binary neutron or QSs via tidal effects. We will show the relation between the tidal Love number ($k_2$) and the tidal deformability ($\lambda$) for the anisotropic QSs in the Starobinsky model. QSs are massive within a limited volume, leading to one of the highly densed objects in our Universe. They form themselves against the tidal deformability $\lambda$ defined as
\begin{eqnarray}
    \lambda = \frac{2}{3} R^5 k_2,
\end{eqnarray}
where $R$ is the radius of the stars. $k_2$ is the dimensionless parameter which is sensitive to EoS and plays a major role to the tidal forces. To obtain the Love number $k_2$, we can solve the following equation written in terms of compactness $C = M/R$
\begin{eqnarray}
    k_2 &=& \frac{8C^5}{5} \left( 1 - 2 C^2 \right) \left( 2 - y_R + 2 C (y_R - 1) \right) \nonumber  \\
    && \times \Bigg( 2 C (6 - 3y_R + 3 C (5Y_R - 8)) \nonumber \\
    && + 4C^3 (13 - 11 y_R + C (3y_R - 2) + 2 C^2 (1+ y_R)) \nonumber \\
    && + 3 (1 - 2C)^2 (2 - y_R + 2 C (y_R - 1)) \ln (1 - 2 C) \Bigg)^{-1}, \nonumber \\
\end{eqnarray}
where $y_R = y(R)$ is at the radius $R$ and is determined by
\begin{eqnarray}
    r \frac{d y(r)}{dr} + y^2(r) + y(r) F(r) + r^2 Q(r) = 0,
\end{eqnarray}
where
\begin{eqnarray}
    F(r) &=& \left( 1 - 4\pi r^2 \left( \rho(r) - p_r(r) \right) \right) \left( 1 - 2C \right)^{-1} \\
    r^2 Q(r) &=& 4 \pi r^2 \frac{\left( 5 \rho(r) + 9 p_r(r) + \frac{\rho(r) + p_r(r)}{\frac{\partial p_r(r)}{\partial \rho(r)}} \right)}{\left( 1 - 2C \right)} \nonumber \\
    && - 6 \left( 1 - 2 C \right)^{-1} - 4C^2 \frac{\left(1 + \frac{4\pi r^2 p_r(r)}{C} \right)^2}{\left( 1 - 2 C\right)^2}.
\end{eqnarray}
We also introduce the dimensionless tidal deformability 
\begin{eqnarray}
    \Lambda = \frac{\lambda }{M^5} = \frac{2}{3} k_2 \left( \frac{1}{C^5}\right)
\end{eqnarray}
The lower of the $\Lambda$ value becomes, the higher of the stars could hold against the tidal deformation.}

\textcolor{black}{
We also plot the dimensionless tidal deformability $\Lambda$ against the total mass $M_{\odot}$ in Fig.~\ref{fig_tidal} for the parameter set defined in Fig.~\ref{fig_vary_a} to \ref{fig_vary_beta}. We also compare our results to the observational result of GW170817 as the constraint on the tidal deformability.} {\color{black}
Anisotropy is expected to imprint observable signatures on the gravitational wave spectrum of QSs. In particular, changes in the tidal deformability parameter $\Lambda$ (as illustrated in  Fig.~\ref{fig_tidal}) can alter the inspiral waveform during binary mergers, while shifts in the oscillation mode frequencies (e.g., f-modes and p-modes, see \cite{Andersson:1998,Kokkotas:1999}) due to increased compactness for positive $\beta$ may lead to higher-frequency post-merger signals.
}

\section{Conclusion}\label{sec5}

In this work, we investigated the structure and stability of QSs within the framework of $R^2$ gravity, specifically using the Starobinsky model, where $f(R) = R + a R^2$. We explored the impact of anisotropic pressures and quark matter equations of state (EoS) with $\mathcal{O}(m_s^4)$ corrections on the mass-radius ($M-R$) and mass-central density ($M-\epsilon_c$) relations for QSs. By solving the modified TOV equations, we examined the role of the $R^2$ gravity parameter $a$, as well as the anisotropy parameters $B_{\perp}$ and $a_4^{\perp}$ derived from the EoS, on the properties and stability of these compact objects.

We found that the presence of anisotropy in the pressure significantly affected the mass and radius of QSs. In particular, our results indicated that increasing the anisotropy parameter $\beta$ enhanced the maximum mass of the QSs, suggesting the possibility of forming super-massive pulsars, consistent with recent astronomical observations, such as PSR J0952-0607 and GW170817. The calculated mass-radius relationships and compactness of the QSs adhered to the Buchdahl limit, ensuring that the solutions remained physically viable. We also confirmed that the inclusion of anisotropic pressures does not violate causality, as the sound speeds in both radial and tangential directions satisfied the condition $0 \leq v_s^2 \leq 1$.

Moreover, we analyzed the dynamical stability of the QSs by computing the adiabatic indices in the radial and tangential directions. Our results showed that the adiabatic indices $\Gamma_r$ and $\Gamma_t$ exceeded the critical value of $4/3$, thereby ensuring the stability of the QSs against radial perturbations. We also applied the static stability criterion, verifying that the gravitational mass increased monotonically with central density, which further supported the stability of the configurations.

In summary, our study provided a detailed analysis of QSs in $R^2$ gravity, incorporating both pressure anisotropy and interacting quark matter EoS. The results demonstrated that QSs in $R^2$ gravity could attain masses greater than $2M_{\odot}$, with stability and causality preserved. These findings contribute to the growing body of evidence supporting the existence of super-massive QSs, as inferred from recent astrophysical data.

While the results of this study are promising, several open questions remain. One potential direction for future research is to explore the impact of rotation on the structure and stability of anisotropic QSs in modified gravity, as rotation could significantly alter the maximum mass and stability conditions. Additionally, studying the behavior of QSs under more complex equations of state, particularly those including color superconducting phases, may provide deeper insights into the internal structure of these compact objects. Finally, the gravitational wave signatures from anisotropic QSs in $R^2$ gravity could be explored further, with the aim of distinguishing between different compact object models through future gravitational wave observations.

\appendix
\section{Detailed Derivation of the Field Equations}\label{app:derivation}

{\color{black}
In this Appendix, we outline the derivation of the field Eqs. (\ref{eq:FieldEq1}-\ref{eq:FieldEq4}) starting from the Einstein frame action
\begin{align}\label{eq:A1}
S &= \frac{1}{16\pi G} \int d^4x\, \sqrt{-\tilde{g}} \left[\tilde{R} - 2\,\tilde{g}^{\mu\nu}\partial_\mu\varphi\,\partial_\nu\varphi - V(\varphi)\right] \notag \\
&+S_M[\psi_i,\,\tilde{g}_{\mu\nu}A(\varphi)^2].
\end{align}
By varying the action with respect to the metric $\tilde{g}_{\mu\nu}$, one obtains the Einstein equations. For the static, spherically symmetric metric given in Eq. \eqref{izmet}, the explicit forms of the components of the Einstein tensor (namely, $\tilde{G}_{tt}$, $\tilde{G}_{rr}$, and $\tilde{G}_{\theta\theta}$) are derived accordingly:
\begin{equation}
\tilde{G}_{tt} = e^{2\Phi}\left[\frac{1}{r^2} - \frac{e^{-2\Lambda}}{r^2}(1-2r\Lambda')\right],
\end{equation}
\begin{equation}
\tilde{G}_{rr} = e^{2\Lambda}\left[\frac{1}{r^2} - \frac{e^{-2\Lambda}}{r^2}(1+2r\Phi')\right],
\end{equation}
\begin{equation}
\tilde{G}_{\theta\theta} = e^{-2\Lambda}r^2\left[\Phi'' + (\Phi')^2 - \Phi'\Lambda' + \frac{\Phi' - \Lambda'}{r}\right].
\end{equation}

The total energy-momentum tensor is constructed as the sum of the scalar field contribution,
\begin{equation}\label{A:Tphi}
T_{\mu\nu}^{\varphi} = \partial_\mu\varphi\,\partial_\nu\varphi - \tilde{g}_{\mu\nu}\left[\frac{1}{2}\tilde{g}^{\alpha\beta}\partial_\alpha\varphi\,\partial_\beta\varphi+\frac{1}{2}V(\varphi)\right],
\end{equation}
and that of the anisotropic fluid, which transforms according to
\begin{equation}\label{A:matters}
\tilde{T}_{\mu\nu}=A(\varphi)^2T_{\mu\nu},
\end{equation}
with the Einstein frame matter variables related as
\begin{equation}\label{A:matters2}
\tilde{\epsilon}=A(\varphi)^4\rho,\quad \tilde{P}=A(\varphi)^4P_r,\quad \tilde{P}_{\perp}=A(\varphi)^4P_{\perp}.
\end{equation}

Furthermore, by varying the action with respect to $\varphi$, one obtains the scalar field equation
\begin{equation}\label{A:phi}
\square\varphi - \frac{1}{4}\frac{dV(\varphi)}{d\varphi} = -4\pi G\,\alpha(\varphi)A^4(\varphi)\tilde{T},
\end{equation}
where the coupling constant is defined as
\begin{equation}\label{A:alpha}
\alpha(\varphi)=\frac{d\ln A(\varphi)}{d\varphi}=-\frac{1}{\sqrt{3}}.
\end{equation}

Finally, using the Bianchi identities in conjunction with the static metric \eqref{izmet}, the conservation law yields the modified TOV equation. In this derivation, the anisotropic term $\frac{2}{r}(P_{\perp}-P_r)$ naturally emerges. For further details, the reader is referred to \cite{Staykov:2014mwa,Yazadjiev:2014cza}.

}

\section*{Acknowledgements}

A.~Pradhan expresses gratitude to the IUCCA in Pune, India, for offering facilities under associateship programs. \.{I}.~S. expresses gratitude to EMU, T\"{U}B\.{I}TAK, ANKOS, and SCOAP3 for their academic and/or financial support. \.{I}.~S. and T. T. also acknowledge COST Actions CA22113, CA21106, and CA23130 for their contributions to networking.


\end{document}